\documentclass[aps,natphysics,superscriptaddress,twocolumn,amsmath,amssymb, balance,notitlepage,bibnotes]{revtex4-2}
\usepackage[utf8]{inputenc}
\usepackage{physics}
\usepackage{rotating}
\usepackage{verbatim}
\usepackage{graphicx,tabularx}
\usepackage[sort&compress]{natbib}
\usepackage[T1]{fontenc}
\usepackage{soul}
\usepackage{dcolumn}
\usepackage{bm}
\usepackage[version=4]{mhchem}
\usepackage{booktabs}
\usepackage{gensymb}
\usepackage{color}
\usepackage[colorlinks=true,urlcolor=blue,linkcolor=magenta,citecolor=magenta]{hyperref}
\usepackage{enumitem}

\allowdisplaybreaks

\begin{document}
\title{Kitaev interaction and proximate higher-order skyrmion crystal \\ in the triangular lattice van der Waals antiferromagnet Ni$\mathbf{I_2}$}
\author{Chaebin Kim}
\thanks{These authors contributed equally to this work.}
\affiliation{Center for Quantum Materials, Seoul National University, Seoul 08826, South Korea}
\affiliation{Department of Physics and Astronomy, Seoul National University, Seoul 08826, Republic of Korea}
\affiliation{School of Physics, Georgia Institute of Technology, Atlanta, Georgia 30332, USA}
\author{Olivia Vilella}
\thanks{These authors contributed equally to this work.}
\affiliation{School of Physics, Georgia Institute of Technology, Atlanta, Georgia 30332, USA}
\author{Youjin Lee}
\thanks{These authors contributed equally to this work.}
\affiliation{Center for Quantum Materials, Seoul National University, Seoul 08826, South Korea}
\affiliation{Department of Physics and Astronomy, Seoul National University, Seoul 08826, Republic of Korea}
\author{Pyeongjae Park}
\affiliation{Center for Quantum Materials, Seoul National University, Seoul 08826, South Korea}
\affiliation{Department of Physics and Astronomy, Seoul National University, Seoul 08826, Republic of Korea}
\author{Yeochan An}
\affiliation{Center for Quantum Materials, Seoul National University, Seoul 08826, South Korea}
\affiliation{Department of Physics and Astronomy, Seoul National University, Seoul 08826, Republic of Korea}
\author{Woonghee Cho}
\affiliation{Center for Quantum Materials, Seoul National University, Seoul 08826, South Korea}
\affiliation{Department of Physics and Astronomy, Seoul National University, Seoul 08826, Republic of Korea}
\author{Matthew B. Stone}
\affiliation{Neutron Scattering Division, Oak Ridge National Laboratory, Oak Ridge, Tennessee 37831, USA}
\author{Alexander I. Kolesnikov}
\affiliation{Neutron Scattering Division, Oak Ridge National Laboratory, Oak Ridge, Tennessee 37831, USA}
\author{Yiqing Hao}
\affiliation{Neutron Scattering Division, Oak Ridge National Laboratory, Oak Ridge, Tennessee 37831, USA}
\author{Shinichiro Asai}
\affiliation{Institute for Solid State Physics, The University of Tokyo, Chiba 277-8581, Japan}
\author{Shinichi Itoh}
\affiliation{Institute of Materials Structure Science, High Energy Accelerator Research Organization, Tsukuba 305-0801, Japan}
\author{Takatsugu Masuda}
\affiliation{Institute for Solid State Physics, The University of Tokyo, Chiba 277-8581, Japan}
\author{Sakib Matin}
\affiliation{Center for Nonlinear Studies, Los Alamos National Laboratory, Los Alamos, NM, 87545, USA}
\affiliation{Theoretical Division, Los Alamos National Laboratory, Los Alamos, NM, 87545, USA}
\author{Sujin Kim}
\affiliation{Department of Chemistry and Nano Science, Ewha Womans University; Seoul 03760, Republic of Korea}
\author{Sung-Jin Kim}
\affiliation{Department of Chemistry and Nano Science, Ewha Womans University; Seoul 03760, Republic of Korea}
\author{Martin Mourigal}
\email{mourigal@gatech.edu}
\affiliation{School of Physics, Georgia Institute of Technology, Atlanta, Georgia 30332, USA}
\author{Je-Geun Park}
\email{jgpark10@snu.ac.kr}
\affiliation{Center for Quantum Materials, Seoul National University, Seoul 08826, South Korea}
\affiliation{Department of Physics and Astronomy, Seoul National University, Seoul 08826, Republic of Korea}
\date{\today}

\begin{abstract}
Topological spin textures, such as magnetic skyrmions, are a spectacular manifestation of magnetic frustration and anisotropy. Most known skyrmion systems are restricted to a topological charge of one, require an external magnetic field for stabilization, and are only reported in a few materials. Here, we investigate the possibility that the Kitaev anisotropic-exchange interaction stabilizes a higher-order skyrmion crystal in the insulating van der Waals magnet $\mathrm{NiI_2}$. We unveil and explain the incommensurate static and dynamic magnetic correlations across three temperature-driven magnetic phases of this compound using neutron scattering measurements, simulations, and modeling. Our parameter optimisation yields a minimal Kitaev-Heisenberg Hamiltonian for $\mathrm{NiI_2}$ which reproduces the experimentally observed magnetic excitations. Monte Carlo simulations for this model predict the emergence of the higher-order skyrmion crystal but neutron diffraction and optical experiments in the candidate intermediate temperature regime are inconclusive. We discuss possible deviations from the Kitaev-Heisenberg model that explains our results and conclude that $\mathrm{NiI_2}$, in addition to multiferroic properties in the bulk and few-layer limits, is a Kitaev bulk material proximate to the finite temperature higher-order skyrmion crystal phase.
\end{abstract}
\maketitle

\section*{Introduction}

Magnetic skyrmions, the topological solitons defined by the winding number of a spin texture \cite{Fert2017, Nagaosa2013}, are a fascinating manifestation of complex magnetism and a possible platform for spintronics applications \cite{Parkin2015}. Many realizations of magnetic skyrmions have been reported following the discovery of a skyrmion crystal – a phase of matter characterized by a periodic arrangement of magnetic skyrmions – in metallic MnSi under an applied magnetic field \cite{Muhlbauer2009, Rler2006}. However, for all reported materials \cite{Nagaosa2013, Wang2021, Gao2020}, the topological number (or topological charge) $N_{\rm sk}$ of the individual solitons is limited to one (skyrmions) or half (merons) \cite{Nagaosa2013, Wang2021, Gao2020}. However, higher-order skyrmions with $N_{\rm sk}$ > 1 are theoretically possible \cite{Ozawa2017}. Crystalline phases of such magnetic objects display extreme spin noncoplanarity and are predicted to host unconventional transport phenomena and the possibility to control their topological number using external perturbations \cite{Ozawa2017, Amoroso2020, Zhang2016}.

Two distinct mechanisms are known to stabilize $N_{\rm sk}$ = 1 skyrmion crystal phases: the Dzyalosinskii-Moriya (DM) interaction in non-centrosymmetric systems or the Ruderman-Kittel-Kasuya-Yoshida (RKKY) interaction in metallic systems \cite{Wang2021, Batista2016, Hirschberger2019, Lin2016, Hayami2016, Hayami2021, Hayami2017}. However, a finite magnetic field is invariably required to stabilize the skyrmion phase. In this context, a path to realizing higher-order skyrmion phases, especially with $N_{\rm sk}$ = 2 (SkX-2 phase), has been theoretically suggested for Kondo lattice models \cite{Ozawa2017, Batista2016, Hayami2017}. A new mechanism was recently discovered to stabilize the SkX-2 phase: the Kitaev exchange interaction \cite{Amoroso2020}. In this case, the bond-directional Ising-like anisotropic exchange interactions select a SkX-2 phase that differs from all the previous realizations \cite{Zhang2016, Yu2014, Leonov2015} because it is stabilized in a thermal equilibrium state without a magnetic field. However, the magnetic field can control the topological charge of individual solitons \cite{Ozawa2017, Amoroso2020}. Although, to our knowledge, there is no experimental report of the SkX-2 phase in a bulk material, our present work shows that the van der Waals multiferroic antiferromagnet $\mathrm{NiI_2}$ is a promising candidate.

\begin{figure*}[t]
    \centering
    \includegraphics[width=1.0\linewidth]{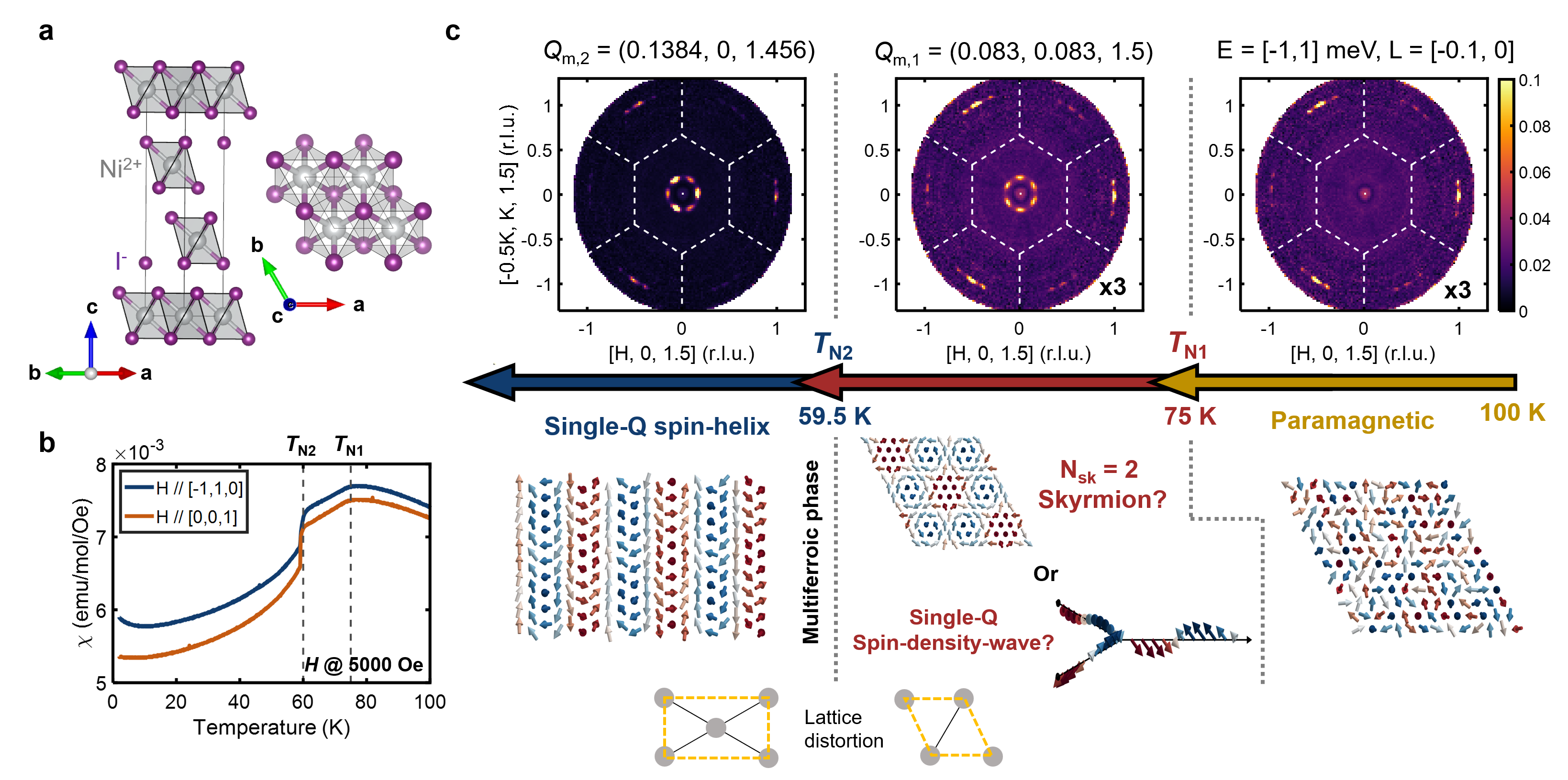}
    \caption{\textbf{Crystal structure and magnetic phases in $\mathbf{NiI_2}$} \textbf{a,} Crystal structure of $\mathrm{NiI_2}$ in $R\bar3m$. The grey spheres represent $\mathrm{Ni^{2+}}$ ions, and the purple spheres represent Iodine ions. \textbf{b,} Magnetic susceptibility of single crystal $\mathrm{NiI_2}$ measured with different directions under a magnetic field $\mu_0H$ = 0.5 Tesla. The two grey lines serve as a guide to the eye, marking two magnetic phase transitions. \textbf{c,} Schematic representation of magnetic phase transition in $\mathrm{NiI_2}$. The upper panels show the magnetic and lattice Bragg peaks of $\mathrm{NiI_2}$ at $T$ = 100 K (right), $T$ = 70 K (Middle) and $T$ = 4 K (left). For elastic neutron scattering data, the integration range is $\Delta L = [-0.1, 0]$ and $\Delta E = [-1, 1]$ meV with an incident neutron energy of $E_i$ = 11.5 meV. The intensity of the $T$ = 100 K and 70 K data was 3 times increased compared to $T$ = 4 K, as shown in the bottom right of the figures. The lower panels depict cartoons of the possible magnetic structure in each magnetic phase. The arrow colour indicates the magnitude of the $z$-direction magnetic moment at each spin site (consistent with fig.\ref{fig:4}). On the bottom of \textbf{c}, cartoon of the lattice transformation at the multiferroic phase transition.}
    \label{fig:1}
\end{figure*}

$\mathrm{NiI_2}$ belongs to a large family of van der Waals (vdW) transition-metal dihalides (see Fig.\ref{fig:1}a for the crystal structure) with rich magnetic properties \cite{McGuire2017}. All triangular-lattice iodine-based compounds in this series display exotic magnetism \cite{Kuindersma1981, Ju2021, Kurumaji2013, Son2022, Song2022}, including multipolar excitations \cite{Bai2021} and Kitaev exchange interactions enhanced by the spin-orbit coupling in the iodine ligands \cite{Kim2023}. However, $\mathrm{NiI_2}$ is unique as the only known multiferroic with a stable multiferroic phase down to the bilayer \cite{Ju2021} and the monolayer limits \cite{Song2022}. In bulk form, $\mathrm{NiI_2}$ displays two magnetic phase transitions at $T_{\rm N1}$ = 75 K and $T_{\rm N2}$ = 59.5 K; see Fig.\ref{fig:1}b for susceptibility measurements and Fig.\ref{fig:1}c for an overview of temperature-dependent elastic neutron scattering patterns and cartoons of the associated magnetic structures. When the sample is cooled through the second magnetic transition at $T_{\rm N2}$, the paramagnetic centrosymmetric space group $R\bar3m$ changes into an orthorhombic structure \cite{Kuindersma1981}. Concomitantly, a multiferroic phase is established with an incommensurate helical magnetic structure (MF-Helix) with the propagation vector $\mathbf{Q}_{m2} = $ (0.1384, 0, 1.457) \cite{Kuindersma1981}, electric polarization ${\bf P}$ along the (1, 1, 0) direction, and a strong optical second harmonic generation (SHG) signal \cite{Ju2021, Son2022, Song2022}. Given the small lattice distortion, all reciprocal lattice vectors are indexed in the paramagnetic cell throughout this text. The intermediate phase between $T_{\rm N1}$ and $T_{\rm N2}$ is less understood than the multiferroic phase and is characterized by a commensurate out-of-plane magnetic structure with propagation vector $\mathbf{Q}_{m1}$ = (0.083, 0.083, 1.5). The absence of the SHG signal suggests that the inversion symmetry is preserved in the intermediate phase, seemingly indicating a simple collinear antiferromagnetic structure \cite{Ju2021, Son2022, Song2022}. However, first-principles calculations for $\mathrm{NiI_2}$ contrast with this simple scenario: these yield sizable frustrated further-neighbor and Kitaev exchange interactions predicted to conspire in stabilizing a SkX-2 phase in the monolayer limit \cite{Ozawa2017, Stavropoulos2019, Li2023}. This leads to the tantalizing possibility of realizing the SkX-2 phase -- or one of its proximate phases -- in $\mathrm{NiI_2}$ and warrants a full experimental investigation of the nature of the intermediate phase. Experimental investigation of spin correlations in bulk  $\mathrm{NiI_2}$ is expected to be difficult due the comparatively small moment, the large energy bandwidth of the system, and the incommensurate and fully three-dimensional nature of the magnetic order.

\section*{Results}

We use elastic and inelastic neutron scattering measurements, supported by semiclassical spin dynamics simulations and inverse modeling techniques, to refine the model Hamiltonian of bulk $\mathrm{NiI_2}$. We first discuss spin excitations in the correlated paramagnet (PM) just above $T_{\rm N1}$. Spin correlations unambiguously reflect the presence of antiferromagnetic Kitaev interactions. By further considering excitations in the intermediate (IN) and multiferroic (MF) phases, we fully quantified the relevant energy scales of the material system and obtained a quantitative understanding of its exchange Hamiltonian. Classical Monte-Carlo and Landau-Lifshitz simulations for this refined model correctly capture the experimental static and dynamic spin correlations in the intermediate phase between $T_{\rm N1}$ and $T_{\rm N2}$. Simulations for the refined parameters favor a multi-$\mathbf{Q}_m$ over a single-$\mathbf{Q}_m$ ground-state, i.e., a SkX-2 magnetic ground state. While our experiments cannot unambiguously distinguish between these two scenarios, our data-informed approach points at the importance of the Kitaev interaction to explain the intermediate temperature phase in bulk $\mathrm{NiI_2}$ what places this material, at, or proximate to, a high topological number skyrmion crystal phase.

\begin{figure*}[t!]
    \centering
    \includegraphics[width=0.9\linewidth]{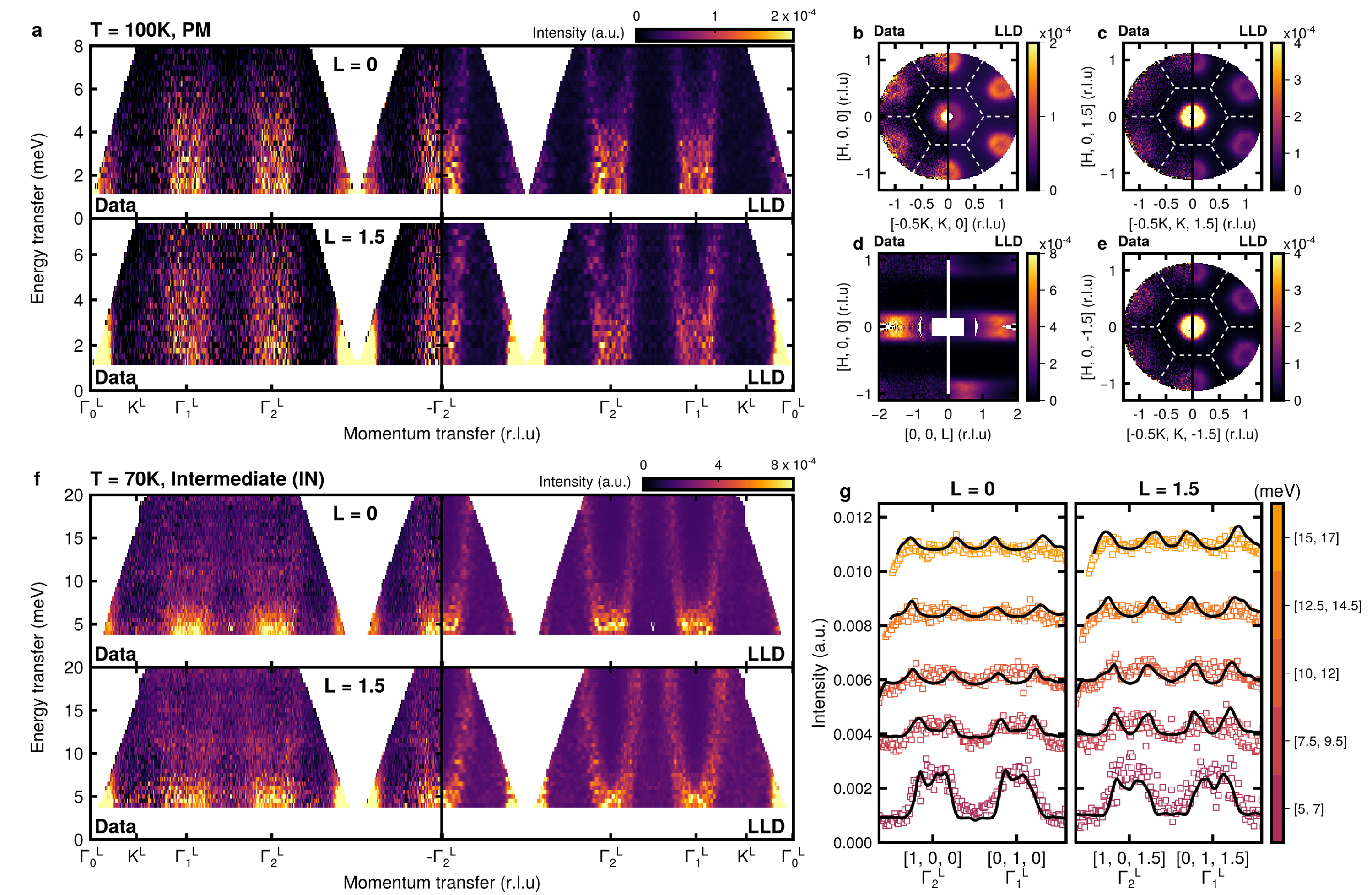}
    \caption{\textbf{Energy-resolved paramagnetic and intermediate-phase excitations in $\mathbf{NiI_2}$} \textbf{a,} Energy-resolved neutron scattering intensity in the paramagnetic regime at $T$ = 100 K along the symmetry directions of the hexagonal Brillouin zone with incident energy of $E_i$ = 11.5 meV (see Fig.\ref{fig:S2} for the directions). The upper figure is for average $\bar{L}$ = 0, and the bottom is for $\bar{L}$ = 1.5. The left side is the data, and the right side is the LLD simulation for our optimized exchange model. For the high symmetry points, $\Gamma_{0}^{L} = (0, 0, \bar{L})$, $\Gamma_{1}^{L} = (0, 1,\bar{L})$, $K^L = (1/3, 1/3, \bar{L})$, and $\Gamma_{\pm n}^{L} = (\pm n, 0, \bar{L})$. Throughout, the intensity is integrated over $\Delta L = \bar{L} + [-0.2, 0.1]$ \textbf{b-e,} Energy-integrated paramagnetic scattering intensity integrated over $\Delta E$ = [1, 6] meV from the $E_i$ = 11.5 meV data at $T$ = 100 K. \textbf{b-c, e} shows the $(H, K, \bar{L})$ plane with $\bar{L} = 0, 1.5, -1.5$, respectively. \textbf{d} shows the $(H, \bar{0}, L)$ plane integrated over $\Delta K = [-0.1, 0.1]$. \textbf{f,} Energy-resolved neutron-scattering intensity at $T$ = 70 K with an incident energy of $E_i$ = 40 meV. The upper figure is $\bar{L}$ = 0, and the bottom is $\bar{L}$ = 1.5. \textbf{g,} shows constant-energy line cut through Fig.\ref{fig:2}f near the two Gamma points, $\Gamma_{2}^{L}$ and $\Gamma_{1}^{L}$, with $\bar{L}$ = 0 and 1.5 at $T$ = 70 K. The colour bar indicates the energy-integrated region of each constant-energy slice. Black lines show the LLD simulation with a given energy integration range. }
    \label{fig:2}
\end{figure*}

We start by defining a minimal bilinear magnetic Hamiltonian for $\mathrm{NiI_2}$ based on the symmetry constraints of the paramagnetic $R\bar3m$ space group, allowing bond-dependent anisotropic exchange interactions. According to first-principles calculations\cite{Ozawa2017, Stavropoulos2019, Li2023}, the Heisenberg and Kitaev interactions are expected to dominate on the nearest-neighbor bonds, and we restrict our analysis to these two terms. Further neighbor interactions are also required to explain our results. However, since most Ni-based compounds comprising networks of edge-sharing octahedral show small second nearest neighbor intralayer coupling $J_2$, we neglect this parameter in our model\cite{Scheie2023, Stock2010, Wildes2022, Gao2021}. Thus, our magnetic Hamiltonian contains four independent parameters and reads:
\begin{eqnarray*}
    \mathcal{H} &=& \sum_{\left<i,j\right>_1\in\{\alpha, \beta, \gamma\}}\left[J_1 \mathbf{S}_i\cdot \mathbf{S}_j + KS_i^\gamma S_j^\gamma\right] + \sum_{\left<i,j\right>_n}^{3, c_1, c_2}J_n\mathbf{S}_i\cdot \mathbf{S}_j
    \label{eq:hamiltonian}
\end{eqnarray*}
where $n$ = 1, 3 indicates the intralayer $n$-th nearest neighbor, $n$ = $c_1$, $c_2$ indicates the interlayer $n$-th nearest neighbor coupling, and $K$ indicates the Kitaev interaction for $X$-,$Y$-,and $Z$-bond (see Fig.\ref{fig:S1} for definitions). 

With a minimal model at hand, we examine the possible zero-temperature magnetic structure of $\mathrm{NiI_2}$ using the direct classical energy minimization and the Luttinger-Tisza method \cite{Litvin1974}. Assuming that a single-$\mathbf{Q}_m$ magnetic structure is stabilized without lattice distortion, we find that the form of $\mathbf{Q}_m$ in the triangular plane only depends on the sign of the Kitaev term (see Supplementary Information and Fig.\ref{fig:S2}). An antiferromagnetic (AFM, $K\!>\!0$) Kitaev term stabilizes a propagation vector $\mathbf{Q}_m(K\!>\!0)$ along the (1,1,0) direction. In contrast, a ferromagnetic term (FM, $K\!<\!0$) yields $\mathbf{Q}_m(K\!<\!0)$ along the (1,0,0) direction. The additional role of the interlayer coupling can be captured by direct energy minimization. The first-neighbor interlayer coupling $J_{\rm c1}$ alone cannot explain the out-of-plane incommensurate order observed below $T_{\rm N2}$. In contrast, the second-neighbor interlayer coupling $J_{c2}$ yields the correct doubly incommensurate $\mathbf{Q}_{m2}$ if the constraint $J_{\rm c1} = -0.031J_{c2}$ is fulfilled. However, the incommensurate nature of out-of-plane magnetic order only emerges when the in-plane order propagates along the (1,0,0) direction. In the experiment, $\mathrm{NiI_2}$ has $\mathbf{Q}_m \parallel (1,1,0)$ at $T_{\rm N2} < T < T_{\rm N1}$ without lattice distortion. Thus, calculations point to an antiferromagnetic Kitaev term consistent with previous theoretical studies \cite{Amoroso2020, Stavropoulos2019, Li2023}.

\begin{figure*}[t!]
    \centering
    \includegraphics[width=0.9\linewidth]{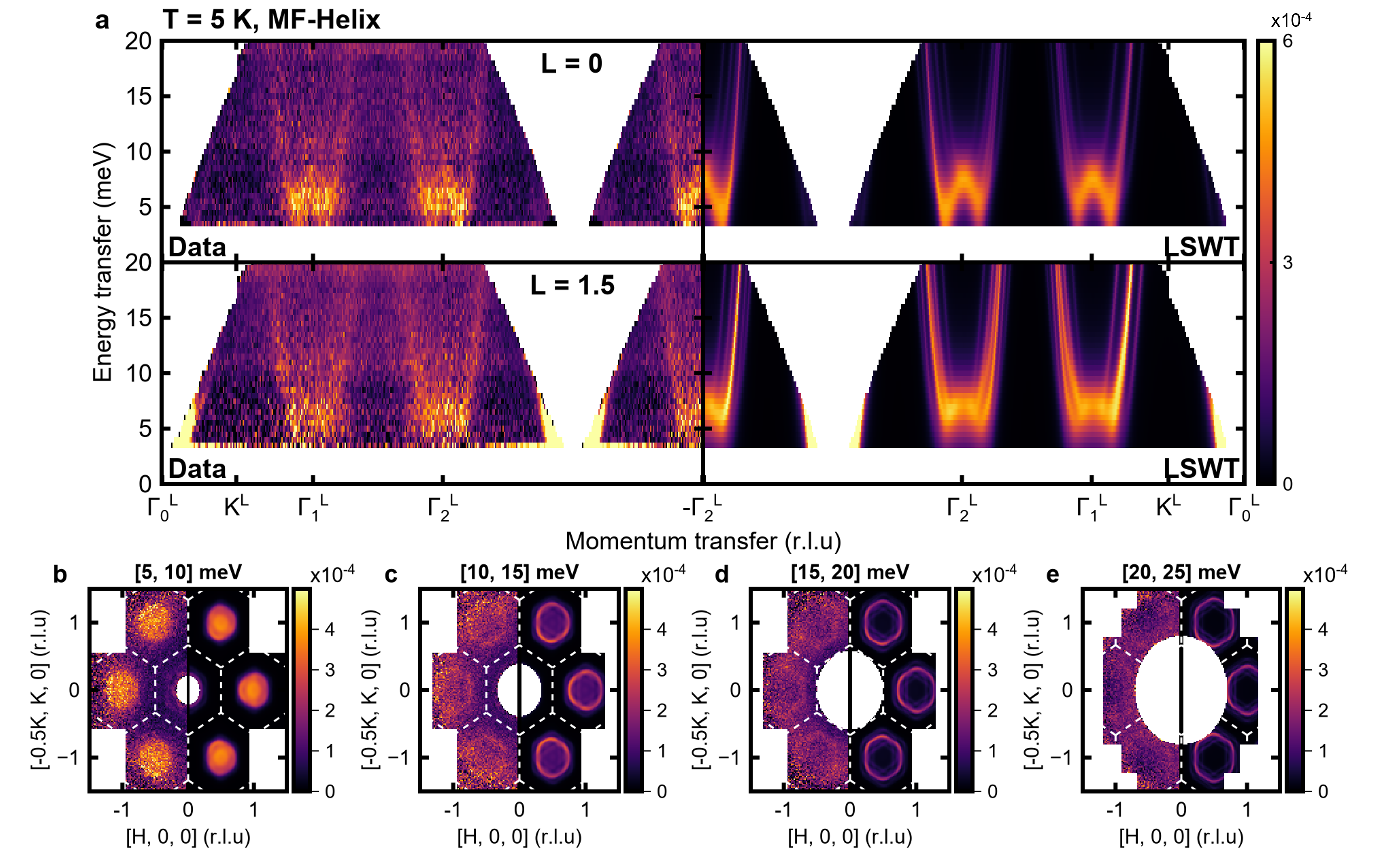}
    \caption{\textbf{Spin-wave spectrum of in the low-temperature phase of $\mathbf{NiI_2}$} \textbf{a,} Spin-wave spectrum of the MF-Helix phase at $T$ = 5 K with an incident energy of $E_i$ = 40 meV. The upper panel shows the $\bar{L}$ = 0 spectrum, and the bottom panel shows the spectrum at $\bar{L}$ = 1.5. The left panels display the experimental data, while the right panel presents simulations based on the linear spin-wave theory (LSWT) using the best-fit parameters for the PM phase. \textbf{b-e,} Energy-integrated spin-wave spectrum integrated over $\Delta L$ = [-0.25, 0.25]. The energy integration range for each figure is specified at the top of the respective panel. The left side of each panel shows experimental data and the right side shows the LSWT simulation. The colour bar represents the intensity of the dynamic structure factor of both data and simulations.}
    \label{fig:3}
\end{figure*}

To gain more insight into microscopic exchange interactions, we turn to the magnetic response in the correlated paramagnetic regime of $\mathrm{NiI_2}$ just above $T_{\rm N1}$. This regime allows the extraction of model parameters without describing spin waves that emerge from a complex, potentially multi-$\mathbf{Q}_m$, magnetic structure, does not require precise knowledge of the instrument resolution function, and assumes that thermal fluctuations dominate over quantum effects in the spectral response \cite{Paddison2020}. Moreover, the paramagnetic signal is sensitive to bond-dependent anisotropy through an out-of-plane intensity dependence originating from the polarization factor of the neutron scattering cross-section \cite{Paddison2020}. The momentum transfer ${\bf Q}$ and energy transfer $E$ dependence of the paramagnetic signal of $\mathrm{NiI_2}$ at $T = 100$ K is shown in Fig. \ref{fig:2}a along several high-symmetry paths in reciprocal space ${\bf Q} = (H, K, \bar{L})$ for an average $\bar{L}$ = 0 and $\bar{L}$ = 1.5 (see Fig.\ref{fig:S3} for the definition of paths in the hexagonal Brillouin zone). The intensity of neutron scattering is concentrated close to the Brillouin zone center due to the long real-space length associated with $2\pi/|{\bf Q}_m|$. In addition, Fig.\ref{fig:2}b-e shows the energy-integrated ($\Delta E = $[1, 6] meV) paramagnetic signal (inelastic diffuse scattering) in two momentum-space planes, ${\bf Q} = (H, K, \bar{L})$ with fixed $\bar{L}$ = 0, 1.5, and ${\bf Q} = (H, \bar{0}, L)$. For $\bar{L}$ = 0, we observe hexagonally-shaped diffuse scattering at low-${\bf Q}$ and near zone centers, Fig.\ref{fig:2}b. When $\bar{L}$ = $\pm 1.5$, the 6-fold intensity becomes 3-fold symmetric with an orientation depending on the sign of $\bar{L}$ (see Fig.\ref{fig:S4}). In our model, this $L$-dependent behavior of the diffuse scattering is the unique signature of a Kitaev interaction \cite{Paddison2020}. 

As a next step, we calculate the neutron scattering intensity at finite temperature using Landau-Lifshitz dynamics (LLD) \cite{Dahlbom2022} for the Hamiltonian of Eq.\ref{eq:hamiltonian} under the constraints imposed by the low-temperature magnetic structure. We used a Bayesian optimization algorithm to estimate the exchange parameters by globally minimizing the $\chi^2$ loss function between the model predictions and the momentum- and energy-dependence of the inelastic paramagnetic scattering at $T$ = 100 and 200 K. The data is split into six representative two-dimensional slices (two momentum-energy slices in Fig.\ref{fig:2}a and Fig.\ref{fig:S5}, four constant-energy slices with $\Delta E$ = [1, 6] meV in Fig.\ref{fig:2}b-e and Fig.\ref{fig:S5}b-e) for each temperature, and the loss function is calculated over all intensity pixels in the set. Because the $T$ = 100 K data only covers a bandwidth of 8 meV, we used the $T$ = 70 K signal to better capture the scale of the exchange interactions by reaching an energy transfer of 20 meV. This is justified by the insensitivity of the high-energy inelastic signal at $T$ = 70 K to details of the underlying magnetic order, given the proximity to $T_{\rm N1}$. Bayesian optimization yields $J_1$ = -7.4(1) meV, $K$ = 2.5(5) meV, $J_3$ = 2.7(1) meV and $J_{c2}$ = 1.2(1) meV as best parameters (see the supplementary information and Fig.\ref{fig:S6} for details). LLD calculations with these parameters agree with the data at $T$ = 100 K, as shown on the right side of Fig.\ref{fig:2}a and the bottom half of Fig.\ref{fig:2}b-e. Fig.\ref{fig:2}f shows the excitations in the intermediate phase at $T$ = 70 K, for which our model faithfully captures the V-shaped excitations near the zone centers. Constant energy slices at $T$ = 70 K, shown in Fig.\ref{fig:2}g, clarify the nature of the V-shaped excitations as dispersing coherent modes as a two-peak structure emerges in both data and simulations.

We now turn to the broadband measurements of the magnetic excitation spectrum in the MF-Helix phase at $T$ = 5 K. Fig.\ref{fig:3}a shows that broad spin-wave like excitations emerge from the $\mathbf{Q}_{m2}$ points of reciprocal space, with a bandwidth greater than the $E$ = 20 meV reach of our measurements and a local maximum around $E$ = 8 meV. The spin gap at the magnetic Bragg peak is less than 0.3 meV (see Fig.\ref{fig:S7}). Since the lattice distorts below $T_{\rm N2}$ \cite{Kuindersma1981}, we performed simulations in the phase using the exchange parameters obtained above but with a slight deviation (1\% of $J_1$) of the nearest-neighbor Heisenberg interaction to mimic the distorted triangular lattice and stabilize the observed single-$Q_m$ magnetic order (see Fig.\ref{fig:S1}). Combining an incommensurate magnetic order with a spin-anisotropic Kitaev interaction makes it impossible to solve the spin-wave Hamiltonian in the conventional, single-${\bf Q}_m$ spiral approximation. Instead, we simulated the spin-wave spectrum in the low-temperature limit by adopting a $1\times14\times2$ super-cell whose dimensions approximate the inverse length of the ${\bf Q}_{m2}$ propagation vector. We minimized the classical energy using the classical Monte-Carlo simulation and performed a linear spin-wave theory for the obtained classical equilibrium spin configuration. While this approach cannot fully capture the incommensurate nature of out-of-plane order and excitations, the simulations reasonably describe the experimental results without further tuning exchange interactions, see Fig.\ref{fig:3}b-e for constant energy slices.

\section*{Discussion}

\begin{figure}[t!]
    \includegraphics[width=1.0\columnwidth]{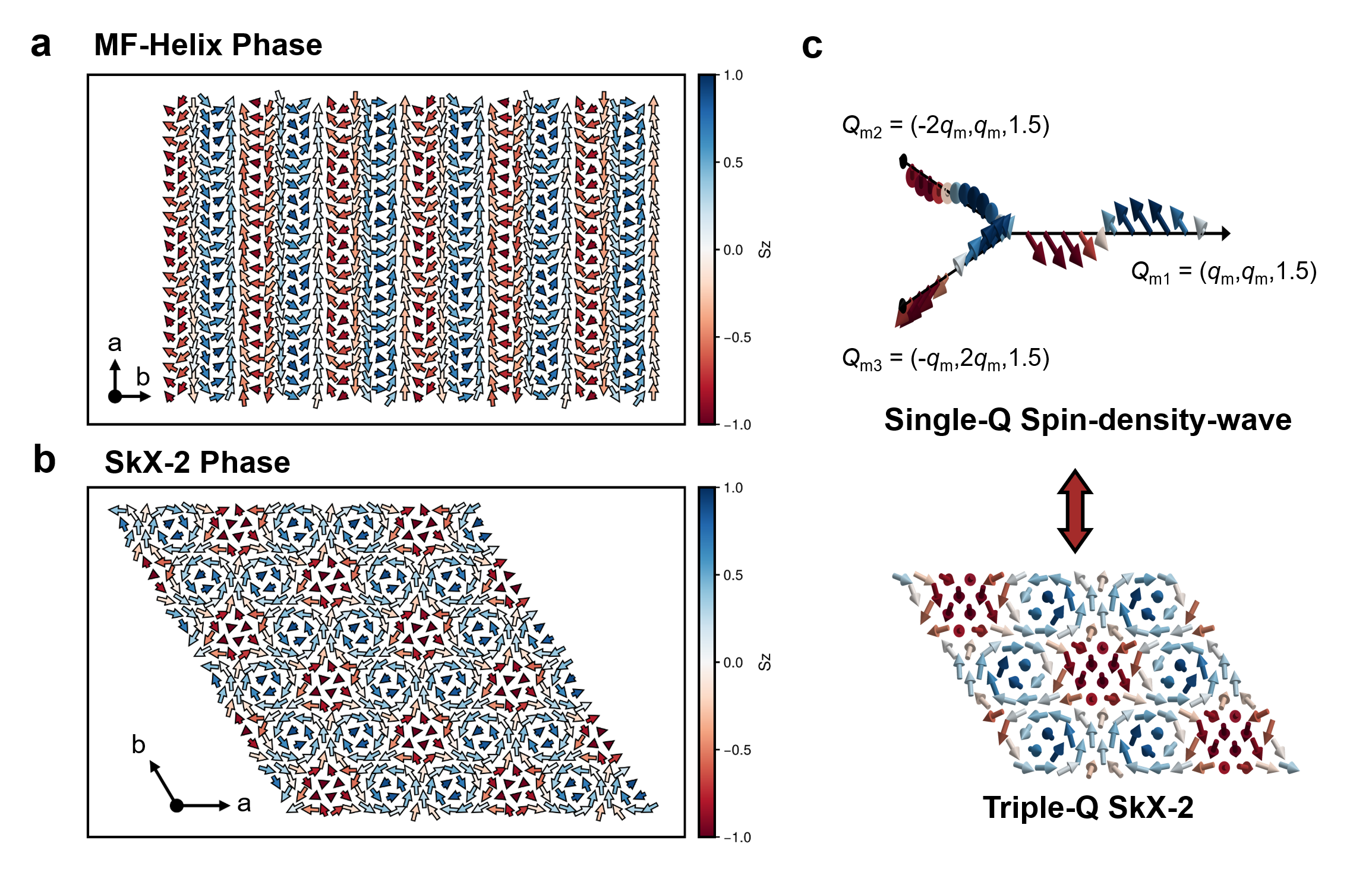}
    \centering
    \caption{\textbf{Lattice-symmetry dependent magnetic ground state in $\mathbf{NiI_2}$} \textbf{a,} Real-space spin configuration of the spiral phase obtained from the Monte Carlo simulation on an orthorhombic lattice. The colour bar indicates the $z$-direction spin moment. \textbf{b,} Real-space spin configuration of the SkX-2 phase based on the classical MC simulation with a hexagonal lattice. \textbf{c,} Schematic representation of the SkX-2 spin texture decomposed into three spin-density-wave modes.}
    \label{fig:4}
\end{figure}

Our investigation establishes a minimal exchange Hamiltonian that captures the elastic, diffuse, and inelastic neutron scattering response of $\mathrm{NiI_2}$ in three distinct thermodynamic phases. With this success, we turn to the precise nature of the magnetic ground state in the intermediate phase between $T_{\rm N1}$ and $T_{\rm N2}$. As neutron scattering experiments average over the entire sample, it is challenging to determine whether the intermediate phase corresponds to a multi-${\bf Q}_m$ structure or domains of a single-${\bf Q}_m$ structure. This distinction is made crisp by Monte-Carlo simulations of the low-temperature magnetic structure with our best model parameters and for two different lattice symmetries (see the Supplementary Information). Remarkably, for the orthorhombic lattice structure that supports the multiferroic phase, the magnetic ground state corresponds to a spin helix with an in-plane propagation vector along the (1,0,0) direction (see Fig.\ref{fig:4}a). However, for the hexagonal lattice structure, the in-plane propagation vector changes to (1,1,0) with antiferromagnetically coupled layers, and the spin texture is highly non-coplanar; this is a clear signature of the stabilization of the SkX-2 phase in the simulations, see Fig.\ref{fig:4}b. These results are consistent with the intertwined nature of the lattice distortion/multiferroic transition and the change of magnetic structure observed in experiments. To go further, we first discuss the mechanism by which the SkX-2 phase is stabilized in simulations before turning to a critical analysis of our data in the intermediate phase.

To better understand the origin of the non-collinear spin texture analytically, we model the putative SkX-2 magnetic structure using a triple-$\mathbf{Q}_m$ model, where each $\mathbf{Q}_n$, $n=1,2,3$ is associated with a longitudinal spin-density wave as follows \cite{Ozawa2017, Hayami2021, Yambe2021, Okubo2012}:
\begin{eqnarray}
    {\bf S}(r_i) = \sum_{n=1}^3 A_n\hat{e}_n \cos(\mathbf{Q}_n\cdot\mathbf{r}_i+\phi_n)
    \label{eq:skx2_decomposition}
\end{eqnarray}
where ${\bf S}(r_i)$ is the spin moment at position $r_i$, and $A_n$ is the contribution of each of three $\mathbf{Q}_n$, $\hat{e}_n$ are three mutually orthogonal unit vectors, $\phi_n$ is the phase factor associated with each of the propagation vectors $\mathbf{Q}_1 = (q_m, q_m, 1.5)$, $\mathbf{Q}_2 = (-2q_m, q_m, 1.5)$, $\mathbf{Q}_3 = (q_m, -2q_m, 1.5)$ with $q_m$ = 0.083 fixed to the experimental value. We fit this analytical form to the spin configurations obtained from MC simulations, with the unit vectors $\hat{e}_n$ aligned with the axes of the Kitaev anisotropy, which gives excellent agreement for $A_n$ = 0.7765 for each $n$ (see the Supplementary Information). In these conditions, the triple-${\bf Q}_m$ structure describes a perfect SkX-2 crystalline phase with a high topological charge per soliton (see Fig.\ref{fig:4}c). 

We now turn to the  distinction between a single-$\mathbf{Q}_m$ ordered structure with three domains and a single-domain triple-$\mathbf{Q}_m$ structure. This is important because several frustrated multiferroic compounds are known to display a two-step ordering: a single-$\mathbf{Q}_m$ spin-density-wave first sets in, and precedes, upon cooling, the onset of a single-$\mathbf{Q}_m$ helical structure with multiferroic properties \cite{Cheong2007, Gordon2018, Kenzelmann2006, Niermann2014}. This phenomenology usually relies on the existence of uniaxial easy-axis anisotropy. However, $\mathrm{NiI_2}$ appears strikingly different from these examples: we expect the large Kitaev interaction to stabilize a triplet of spin-density-wave-like modes, each polarized along three mutually orthogonal easy axes. It is not energetically favorable to stabilize the single-$\mathbf{Q}_m$ spin-density-wave phase in our hexagonal four parameters model of $\mathrm{NiI_2}$. More generally, the symmetry-allowed exchanges anisotropies for the paramagnetic space group of $\mathrm{NiI_2}$ allow for a deviation from a pure Heisenberg-Kitaev model. We expect these deviations, that are not necessary to explain our current neutron scattering data, to complicate the stabilization of three longitudinal spin-density waves with equal amplitude (the Sk-2 phase). For this reason, it is reasonable to conclude that our modeling effort supports $\mathrm{NiI_2}$ as being proximate to a finite-temperature Sk-2 phase but not necessarily exactly in that phase.

On the experimental side, as inversion symmetry is not broken by the superposition of three spin-density wave modes with equal amplitude, neither SHG nor ferroelectricity is expected in the putative SkX-2 phase, as observed experimentally for $\mathrm{NiI_2}$~\cite{Kurumaji2013}. However, the bulk crystals of $\mathrm{NiI_2}$ display weak linear optical dichroism below ${T_{N1}}$ \cite{Song2022}, indicating that the underlying magnetic structure breaks the 3-fold rotation symmetry of the paramagnetic lattice. Such a finite linear dichroism signal can be interpreted in two ways: an unequal distribution of three single-${\bf Q}_m$ spin-density-wave domains across the sample, or a non-equilateral triple-${\bf Q}_m$ SkX-2 phase with imbalanced $A_n$ amplitudes due to the thermal fluctuations, or sub-leading anisotropic terms in the Hamiltonian as discussed above. Other optical measurements, such as circular dichroism, are complicated by the antiferromagnetic out-of-plane correlation and are inconclusive. 

In an attempt to create a population imbalance between putative single-${\bf Q}_m$ domains, we also performed a neutron diffraction experiment in an applied magnetic field along the $a$-axis. The sample was cooled in zero field from its paramagnetic state at $T=100$~K to $T=62$~K (below $T_{\rm N1}$) and then to $T=5$~K (below $T_{\rm N2}$). The experiment was repeated while field cooling in $\mu_0H=5$~T to $T=62$~K and then $T=5$~K. While the integrated intensity of the $\mathbf{Q}_{m1}$ Bragg peaks (see Fig.~\ref{fig:S9}d) at $T=62$~K does not change appreciably between field-cooled and zero-field cooled conditions, the intensity of the $\mathbf{Q}_{m2}$ Bragg peaks at $T=5$~K significantly redistributes, see Fig.\ref{fig:S9}b. While this behavior is consistent with the existence of a triple-${\bf Q}_m$ magnetic structure below $T_{\rm N1}$, the difference in behavior between the $\mathbf{Q}_{m1}$ and $\mathbf{Q}_{m2}$ peaks could also originate from the elevated temperature and too low magnetic-field scale, and the result of this experiment is inconclusive.

In conclusion, $\mathrm{NiI_2}$ is a unique quantum material that bridges the realm of fundamental Kitaev quantum magnetism with applied two-dimensional thin-layer multiferroicity. Our neutron scattering and semi-classical modeling results demonstrate a sizable Kitaev interaction in this van der Waals triangular-lattice antiferromagnet. Along with four other exchange parameters, this leads to a minimal model that successfully describes the static and dynamical magnetic properties in reciprocal space for three distinct thermodynamic phases: a cooperative paramagnet above 75 K, an intermediate phase between 59.5 and 75 K, and multiferroic helicoidal phase below 59.5 K. In simulations, this model unambiguously stabilizes an intermediate triple-${\bf Q}_m$ structure corresponding to a higher-order SkX-2 skyrmion crystal. Remarkably, this phase corresponds to the coherent superposition of three orthogonal polarized spin-density wave modes and is entirely stabilized by the Kitaev interaction without requiring the two otherwise conventional skyrmion mechanisms of the RKKY and DM interactions. In experiments, the single-${\bf Q}_m$  or triple-${\bf Q}_m$  nature of this intermediate phase remains to be fully understood, given the extreme sensitivity of the latter to slight lattice distortions. This issue can only be address by new, although rather difficult, measurements done in real space, or by few-layer studies, which will remove the out-of-plane magnetic order.

\section*{Methods}
\subsection*{Sample growth and characterization}
Ni$\mathrm{I_2}$ single crystals were synthesized using the chemical vapour transport (CVT) method. Nickel powder and crystalline iodine (99.99\% purity) were weighed in the stoichiometric ratio, with a 5\% excess of iodine, to ensure a complete reaction. The mixture was sealed in a quartz tube under vacuum conditions. The sealed quartz was placed in a horizontal two-zone furnace, where the end zones were heated to 750 and 720 ºC, respectively, over 6 hours. The furnace was held at these fixed temperatures for one week to facilitate crystal growth, followed by slow cooling to room temperature for five days. The resulting crystals formed shiny grey flakes with dimensions of approximately $5 \times 5 \times 0.1 \; mm^3$. The quality of the synthesized crystal was assessed using the magnetic susceptibility measurement. Temperature- and magnetic field-dependent magnetization were measured using a SQUID magnetometer (Quantum Design, MPMS3). Measurements were performed on single crystals across a temperature range of 2 to 300 K, with the magnetic field $H$ = 5000 Oe. 

\subsection*{Neutron scattering measurements}
Inelastic neutron scattering measurements were conducted using the SEQUOIA spectrometer at the Spallation Neutron Source (SNS), Oak Ridge National Laboratory (ORNL), USA \cite{Granroth2010}. For these measurements, 20 pieces of $\mathrm{NiI_2}$ single crystals were coaligned with a total mass of around 1 gram. Because of the hygroscopic nature of the sample, the crystals were covered with hydrogen-free glue (CYTOP) and mounted in an air-tight sample holder attached to the bottom of a closed-cycle refrigerator. To cover the bandwidth of the magnetic excitations, three incident energies ($E_i$ = 8, 11.5, and 40 meV) were used. The sample was measured at various temperatures: $T$ = 5, 70, 100, and 200 K for $E_i$ = 11.5 meV and $T$ = 5 and 70 K for $E_i$ = 40 meV. Data reduction was performed using the MANTID \cite{Arnold2014} software at the ORNL analysis clusters. For the final analysis, the data were symmetrized along the threefold rotational symmetry axis $(0, 0, L)$ of the scattering pattern using the HORACE software package \cite{Ewings2016} to improve statistical reliability. 

Another inelastic neutron scattering measurement was performed using the HRC spectrometer at the MLF, J-PARC, Japan \cite{Itoh2011}. For this experiment, approximately 70 coaligned $\mathrm{NiI_2}$ single crystals with a total mass of 2.5 grams were used. The HRC beamline employed a Fermi chopper (A) operated at 600 Hz, with two incident neutron energies ($E_i$ = 60 and 18 meV). Magnetic Bragg peaks were tracked by measuring at three angles ($\phi$ = -77.5º, -68.5º, and -59.5º) while increasing the temperature from 20 to 80 K.

Neutron diffraction was performed on a single crystal of $\mathrm{NiI_2}$ using the DEMAND~\cite{cryst9010005} diffractometer at the High Flux Isotope Reactor (HFIR), Oak Ridge National Laboratory (ORNL), USA. DEMAND was operated with the wavelength of $\lambda = 1.533$ \AA$^{-1}$. The sample was inserted in a cryomagnet with a maximum field of $\mu_0H=5$~T applied along the $a$ direction. The sample was rotated along the magnetic field axis and the position sensitive detector adjusted to capture up to five equivalent magnetic Bragg peaks in each of the ordered phases of $\mathrm{NiI_2}$. 

\subsection*{Spin dynamics simulation}
The momentum- and energy-dependent neutron scattering intensity in the paramagnetic regime was calculated using the Landau-Lifshitz dynamics (LLD) method using the Sunny.jl code \cite{sunny, Sunny2025}. The Monte Carlo simulation employed a supercell size of $30 \times 30 \times 10$. At each temperature, the initial spin configuration was thermalized using the standard Langevin sampler with a time step of $dt = 0.00135 meV^{-1}$ and the damping parameter $\lambda$ = 0.1 for the Langevin sampler. The thermalization process involved 3000 Langevin time steps to ensure equilibrium. After thermalization, the equation of motion for the sampled spin configuration was integrated using Landau-Lifshitz dynamics. Five spin configurations were sampled, with 1000 Langevin time steps between each sample. The sampled configurations were Fourier-transformed to reconstruct the intensity of the momentum- and energy-dependent neutron scattering. To account for the differences between classical and quantum mechanical behaviour, the calculated classical correlations were adjusted using the classical-quantum correction factor, ensuring accurate comparison with the experimental data. 

We performed LSWT calculations using Sunny.jl \cite{sunny, Sunny2025}. The incommensurate magnetic order combined with the spin-anisotropic Kitaev interaction makes it challenging to solve the spin-wave Hamiltonian using the conventional, single-Q spiral approximation. To overcome this limitation, we adopted the kernel polynomial method \cite{Harry2025} to simulate the low-temperature spectrum. The calculations were conducted with a $1\times14\times2$ supercell chosen to approximate the inverse length of the $Q_{m2}$ propagation vector. We minimized the classical energy using the classical Monte Carlo simulation and performed linear spin-wave theory for the obtained classical equilibrium spin configuration.

\begin{acknowledgements}
The authors thank Cristian Batista and Jong-Seok Lee for their fruitful discussion and constructive comments. The authors also thank Maxim Avdeev for his help with powder neutron diffraction measurement at ECHIDNA, ANSTO. The work at SNU was supported by the Leading Researcher Program of the National Research Foundation of Korea (Grant No. 2020R1A3B2079357) and the National Research Foundation of Korea (Grant No RS-2020-NR049405). The work of O.V. and M.M was supported by the U.S. National Science Foundation through Grant No. NSF-DMR-1750186 and NSF-DMR-2309083. S. Matin acknowledges support from the Center for Nonlinear Studies at Los Alamos National Laboratory. The work of S.K. and S.J.K. was supported by the Pioneer Research Center Program through the National Research Foundation of Korea, funded by the Ministry of Science, ICT \& Future Planning (NRF-2022M3C1A3091988). Part of this research was conducted at the High-Flux Isotope Reactor and Spallation Neutron Source, a DOE Office of Science User Facility operated by the Oak Ridge National Laboratory. The beam time was allocated to SEQUOIA on proposal numbers IPTS-27591 and 29888. The beam time was allocated to DEMAND on proposal numbers IPTS-33254. Another neutron experiment was performed at the Materials and Life Science Experimental Facility of the J-PARC Center under a user program 2022BU1201. One of the authors (J.G.P.) acknowledges the hospitality of the Indian Institute of Science, where the manuscript was finalised, and the financial support of the Infosys Foundation.
\end{acknowledgements}

\section*{Author Contributions}
J.G.P initiated and supervised the project. Y.L., S.K., and S.J.K grew the single crystal. C.K. aligned the sample for measurements. C.K., O.V, P.P., Y.A., W.C, M.B.S, A.I.K, S.I, S.A., T.M., M.M, and J.G.P performed the INS measurement. C.K., Y.H. and M.M performed the single-crystal neutron diffraction measurement. C.K. analyzed the data and performed Landau-Lifshitz dynamics and linear spin-wave calculations. S. M. contributed to the data fitting using the Bayesian optimization. C.K., M.M, and J.G.P. wrote the manuscript with input from all co-authors.

\section*{Competing interests}
The authors declare no competing interests.

\section*{Data availability}
The data that supports the findings of this study are available from the corresponding authors upon reasonable request.

\section*{Codes availability}
The codes that support the findings of this study are available from the corresponding authors upon reasonable request.

\bibliography{nii2}

\clearpage
\widetext
\setcounter{equation}{0}
\setcounter{figure}{0}
\setcounter{table}{0}
\setcounter{page}{1}
\setcounter{section}{0}

\renewcommand{\theequation}{S\arabic{equation}}
\renewcommand{\thefigure}{S\arabic{figure}}
\renewcommand{\thetable}{S\arabic{table}}
\begin{center}
\textbf{\large \scshape Supplemental Information}
\end{center}

\section{Luttinger-Tisza method and classical energy minimization}

The magnetic phase diagram for $\mathrm{NiI_2}$ within the $J_1-K-J_3$ model was calculated using the standard Luttinger-Tisza method \cite{Litvin1974}. Using the Fourier transform, we can rewrite the model Hamiltonian (Eq.\ref{eq:hamiltonian}) as 
\begin{eqnarray*}
    H \, && = \sum_{\rm Q}\mathbf{\Psi}_{\rm -Q}H_{\rm Q}\mathbf{\Psi}_{\rm Q} \\
    \mathrm{where \quad} H_{\rm Q} \, && = \begin{pmatrix} 
    A-B & C & D \\
    C & A+B & E \\
    D & E & A 
    \end{pmatrix} \\
    A \, && = \left[J_{\rm 1}-\frac{K}{3}\right]F_{\rm 1}+J_{\rm 2}F_{\rm 2}+J_{\rm 3}F_{\rm 3} \\
    B \, && = \frac{2}{3}K\left[cos(2\pi Q_{\rm h})-cos(2\pi Q_{\rm k})-cos(2\pi(Q_{\rm h}+Q_{\rm k})\right] \\
    C \, && = -\frac{\sqrt{3}}{3}K\left[cos(2\pi(Q_{\rm h}+Q_{\rm k}))-cos(2\pi Q_{\rm k})\right] \\
    D \, && = -\frac{\sqrt{6}}{3}K\left[cos(2\pi(Q_{\rm h}+Q_{\rm k}))-cos(2\pi Q_{\rm k})\right] \\
    F_{\rm 1} \, && = 2cos(2\pi Q_{\rm h}) + 2cos(2\pi Q_{\rm k}) + 2cos(2\pi(Q_{\rm h} + Q_{\rm k})) \\
    F_{\rm 2} \, && = 2cos(2\pi(2Q_{\rm h}+Q_{\rm k}))+2cos(2\pi(2Q_{\rm k}+Q_{\rm h}))+2cos(2\pi(Q_{\rm h}-Q_{\rm k})) \\
    F_{\rm 3} \, && = 2cos(4\pi Q_{\rm h}) + 2cos(4\pi Q_{\rm k}) + 2cos(4\pi(Q_{\rm h}+Q_{\rm k}))
    \label{eq:LTHamiltonian}
\end{eqnarray*}

where the vector $\Psi_Q = \left\{S_{\mathbf{Q},x}, S_{\mathbf{Q},y}, S_{\mathbf{Q},z}\right\}$ is defined from the Fourier-transformed components $S_{r,a} = \sum_QS_{Q,a}e^{iQ\cdot r}$. Using the Luttinger-Tisza formalism, the ground state is determined by solving the eigenvalue problem $H_Q\Psi_Q = \epsilon_Q\Psi_Q$ while satisfying the weak constraint $\sum_i|S_i|^2 = NS^2$ ($N$ is a number of the lattice sites). We found that the direction of the propagation vector of the $J_1-K-J_3$ model solely depends on the Kitaev interaction if $J_3$ is large enough. An antiferromagnetic (AFM, K > 0) Kitaev term stabilizes a propagation vector $\mathbf{Q}_m(K>0)$ along the (1,1,0) direction. In contrast, a ferromagnetic term (FM, K < 0) yields $\mathbf{Q}_m(K<0)$ along the (1,0,0) direction.

To observe the influence of interlayer coupling, we analytically calculated modulations in the propagation vector along the $c$-axis. Considering the 1st and $2^{nd}$ n.n. interlayer coupling (see Fig.\ref{fig:1}c) and in-plane propagation vector parallel to (1, 0, 0) direction, we can obtain the equation of modulation of the propagation vector along the $c$-axis as
\begin{eqnarray}
    tan\left(\frac{2\pi Q_{\rm l}}{3}\right) = \frac{J_{\rm c2}sin\left(\frac{8\pi Q_{\rm h}}{3}\right)-(J_{\rm c1}+2J_{\rm c2})sin\left(\frac{4\pi Q_{\rm h}}{3}\right)+2J_{\rm c1}sin\left(\frac{2\pi Q_{\rm h}}{3}\right)}{J_{c2}sin\left(\frac{8\pi Q_{\rm h}}{3}\right)+(J_{\rm c1}+2J_{\rm c2})sin\left(\frac{4\pi Q_{\rm h}}{3}\right)+2J_{\rm c1}sin\left(\frac{2\pi Q_{\rm h}}{3}\right)}
    \label{eq:skx2_decomposition}
\end{eqnarray}
Since we have $Q_{\rm l} = 1.457$ $\&$ $Q_{\rm h} = 0.1358$ at the MF-Helix phase, this equation gives $J_{\rm c1} = -0.035J_{\rm c2}$. This implies that $2^{nd}$ n.n. interlayer coupling determines the pitch of the out-of-plane incommensurate order, and $1^{st}$ n.n. interlayer coupling is negligible. This trend is also consistent with the previous DFT calculation \cite{Li2023}.
If the in-plane propagation vector is parallel to the [1, 1, 0] direction, we can obtain an equation of modulation of the out-of-plane propagation vector as
\begin{eqnarray}
    sin\left(\frac{2\pi Q_l}{3}\right)\left[J_{\rm c1}\left(2cos(2\pi Q_{\rm h})+1\right)+J_{c2}\left(2cos(4\pi Q_{\rm h})+1\right)\right] = 0
    \label{eq:skx2_decomposition}
\end{eqnarray}
This equation shows that if the in-plane propagation vector is along the [1, 1, 0] direction, the out-of-plane has no incommensurate order. This result is consistent with our observation at the IN phase, where the out-of-plane propagation vector is commensurate at $Q_{\rm l} = 1.5$.

\section{Fitting process details}
The paramagnetic inelastic neutron scattering intensity was analyzed by calculating using Landau-Lifshitz dynamics (LLD) for classical dipoles as implemented in the Sunny.jl package \cite{sunny, Sunny2025}. The Bayesian optimization method \cite{Nogueira2014, Frazier2018} was employed to extract a model of exchange interactions from the data. We investigated a hyperparameter space with four parameters: $J_{\rm 1}$, $K$, $J_{\rm 3}$ and $J_{\rm c2}$. The searching space was bounded for each parameter with $J_{\rm 1}$ = [-8, -4] meV, $K$ = [0, 8] meV, $J_{\rm 3}$ = [0, 4] meV, and $J_{\rm c2}$ = [0, 4] meV after the initial rough analysis. The final optimization proceeds as follows: 
From the initial choice of parameters, we first performed simulated annealing of the spin structure in a $14 \times 14 \times 6$ crystallographic supercell using a Langevin sampler thermalized from 200 to 0.01 K. For each temperature, we performed 5000 Langevin time steps for thermalization with a time step of $dt = 0.00128 meV^{-1}$ and a damping parameter $\lambda$ = 0.01 for the Langevin sampler. After the thermalization, 4000 Langevin time steps were used between each sample. Each run's calculated heat capacity determined the system's $T_{N,cal}$. From the extracted $T_{N,cal}$, we scaled the temperature for LLD to the observed $T_N$ as follows:
\begin{eqnarray}
    T_{\rm scale} = \frac{T_{\rm data}}{T_{\rm N}}\times T_{\rm N, cal}
    \label{eq:skx2_decomposition}
\end{eqnarray}
where $T_{scale}$ is the temperature used for simulating the LLD with given data with temperature $T_{data}$. We used three different experimental temperature data for Bayesian: $T$ = 70, 100, and 200 K. The fitted data is split into six representative two-dimensional slices (two momentum-energy slices in Fig.\ref{fig:2}a and Fig.\ref{fig:S5}, four constant energy slices with $\Delta E$ = [1, 6] in Fig.\ref{fig:2}b-e and Fig.\ref{fig:S5}b-e) for each temperature, and the loss function calculated over all intensity pixels in the set. Because the $T$ = 100 K data only covers a bandwidth of 8 meV, we used the $T$ = 70 K signal to better capture the scale of the exchange interactions by reaching an energy transfer of 20 meV. This is justified by the insensitivity of the high-energy inelastic signal at $T$ = 70 K to details of the underlying magnetic order, given the proximity to $T_{\rm N1}$. 

\begin{figure}[h]
    \includegraphics[width=1.0\columnwidth]{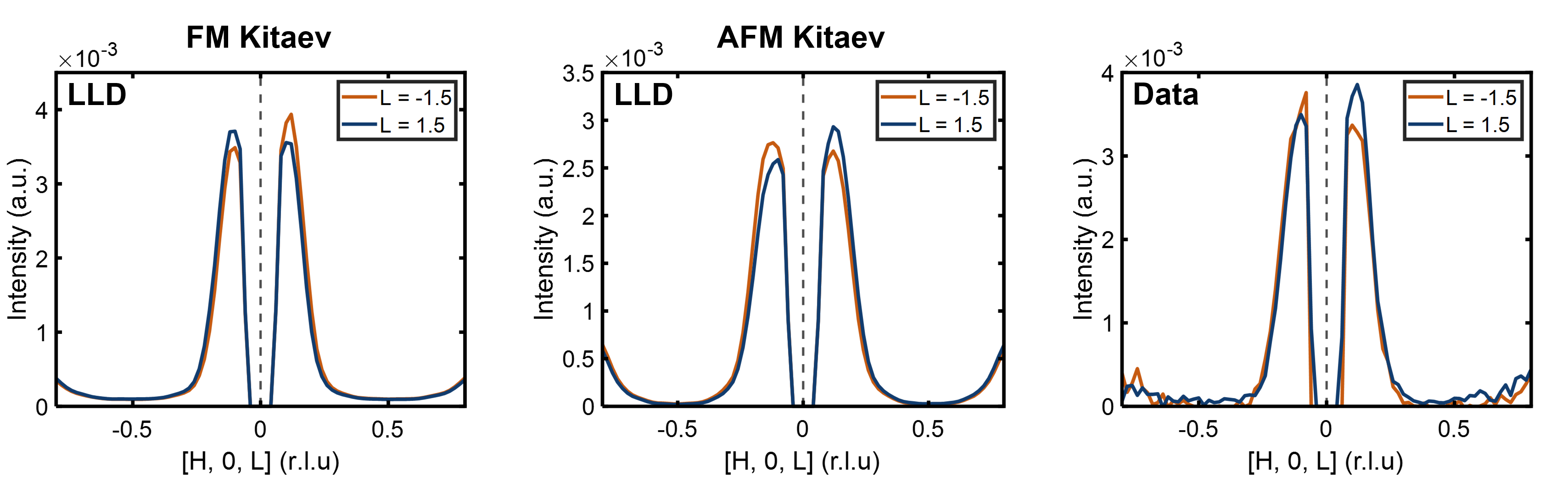}
    \centering
    \caption{The out-of-plane signature of the paramagnetic scattering intensity depends on the sign of the Kitaev interaction. For the FM Kitaev interaction, we used the same parameter set as the AFM Kitaev model and changed the sign of the Kitaev interaction. The paramagnetic scattering intensity was integrated with energy $\Delta E$ = [1, 6] meV and along $\Delta K$ = [-0.015, 0.015] and $\Delta L$ = $±1.5$ + [-0.2, 0.2].}
    \label{fig:S4}
\end{figure}

The momentum- and energy-dependent neutron scattering intensity in the paramagnetic regime was calculated using the Landau-Lifshitz dynamics (LLD) method. The supercell size for the Monte-Carlo simulation was set for $30 \times 30 \times 10$. Before sampling, the initial spin configuration at a given temperature was thermalized using the standard Langevin sampler. We used the time step $dt = 0.00135 \,\, meV^{-1}$ and the damping parameter $\lambda$ = 0.1 for the Langevin sampler. The thermalization was done with the 3000 Langevin time steps. After thermalization, the equation of motion of the sampled spin configuration was integrated using Landau-Lifshitz dynamics. Up to five sampled configurations are Fourier-transformed to reconstruct the momentum- and energy-dependent neutron scattering, with 1000 Langevin time steps between the samples. The calculated classical correlations were corrected using the classical-quantum correction factor to fit the data.
The goodness-of-fit between data and simulations was determined by calculating $\chi^2$:
\begin{eqnarray}
    \chi^2 = \sum_{n = 1}^{14}\sum_{i,j}\left[I_{\rm data, n}(i,j)-\left(S_{\rm 2}+S_{\rm 1}*I_{\rm cal,n}(i,j)\right)\right]^2
    \label{eq:skx2_decomposition}
\end{eqnarray}
where $n$ indicates the number of 2D data slices included in the fit, and $I_{\rm data,n}(i,j)$ shows the intensity of the data at point $(i,j)$ of a 2D slice. $I_{\rm cal,n}(i,j)$ is the intensity of the simulation at $(i,j)$ coordination, and $S_{\rm 1}$, $S_{\rm 2}$ are the scale and background parameters, respectively, to minimize the $\chi^2$ \cite{Proffen1997}. After calculating the goodness-of-fit at one step of the optimization process, the parameters set for the next calculation are guessed based on the Gaussian process model. Fig.\ref{fig:S4} shows the convergence of each parameter based on this Bayesian optimization method. 

\section{Classical Monte Carlo simulation}
The standard Monte Carlo Metropolis simulations with simulated annealing were used for the ground state energy simulation. The supercell size for the Monte Carlo simulation was set for $24 \times 24 \times 2$. For each temperature, we perform 15000 spin flip steps. The temperature step started at 200 K, and the next temperature was 4\% lower than the previous temperature steps. The annealing stopped when the temperature reached 0.01 K. 

\section{The transformation between the cubic axis and the crystallographic axis}
In the Kitaev model, it is conventional to use the cubic axis, whose axes are parallel to the direction of the transition metal and the three upper ligands in the ideal octahedrons. However, given the symmetry in a crystal, it is convenient and realistic to use the crystallographic axes to represent bond-dependent anisotropy \cite{Maksimov2019}. We can convert the generalized Kitaev-Heisenberg model (i.e., $J-K-\Gamma-\Gamma'$ model) into the $J_{\rm 1}-\Delta_{\rm 1}-J_{\rm \pm\pm}-J_{\rm z\pm}$ model in the crystallographic axes by rotating the reference frame. The transformation from the cubic axes to the crystallographic frame $S_{\rm cryst} = \hat{R}S_{\rm cubic}$ is given by the rotation matrix:
\begin{eqnarray}
    \hat{R}_{\rm c} = \begin{pmatrix}
        \frac{1}{\sqrt{2}} & \frac{1}{\sqrt{6}} & \frac{1}{\sqrt{3}} \\
        -\frac{1}{\sqrt{2}} & \frac{1}{\sqrt{6}} & \frac{1}{\sqrt{3}} \\
        0 & -\sqrt{\frac{2}{3}} & \frac{1}{\sqrt{3}}
    \end{pmatrix}
    \label{eq:skx2_decomposition}
\end{eqnarray}
After the rotation of the reference frame, the relation between $J_1-K-\Gamma-\Gamma'$ to $J_1-\Delta_1-J_{\pm\pm}-J_{z\pm}$ can be written as
\begin{eqnarray}
    J_{\rm 1} \, &&= \frac{1}{3}(2J_{\rm 01}+\Delta{\rm 1}J_{\rm 01}+2J_{\rm \pm\pm}+2\sqrt{2}J_{\rm z\pm}) \\
    K \, && = -2J_{\rm \pm\pm}-\sqrt{2}J_{\rm z\pm} \\
    \Gamma \, && = \frac{1}{3}(-J_{\rm 01}+\Delta_{\rm 1}J_{\rm 01}-4J_{\rm \pm\pm}+\sqrt{2}J_{\rm z\pm}) \\
    \Gamma' \, && = \frac{1}{6}(-2J_{\rm 01}+2\Delta_{\rm 1}J_{\rm 01}+4J_{\rm \pm\pm}-\sqrt{2}J_{\rm z\pm}) 
    \label{eq:skx2_decomposition}
\end{eqnarray}
For the Heisenberg-Kitaev model, it can be simplified as follows:
\begin{eqnarray}
    J_{\rm 01} \, && = J_{\rm 1}+\frac{K}{3} \\
    \Delta_{\rm 1} \, && = 1 \\
    J_{\rm \pm\pm} \, && = \frac{J_{\rm z\pm}}{2\sqrt{2}}=-\frac{K}{6}
    \label{eq:skx2_decomposition}
\end{eqnarray}

\section{Analytic form of the Skyrmion phase}
We can write the $N_{sk}=2$ Skyrmion as an approximation of a linear combination of three spin-density-wave as follows \cite{Yambe2021, Ozawa2017}:
\begin{eqnarray}
    \mathbf{S}(\mathbf{r}_{\rm i}) \simeq \sum_{n=1}^3A_{\rm n}\hat{e}_{\rm n}cos\left(\mathbf{Q}_{\rm n}\cdot\mathbf{r}_{\rm i}+\phi_{\rm n}\right)
    \label{eq:skx2_decomposition}
\end{eqnarray}
where $\mathbf{S}(\mathbf{r}_{\rm i})$ is the spin moment at position $\mathbf{r}_{\rm i}$ and $A_{\rm n}$ is the ratio, $\hat{e}_{\rm n}$ is the unit vector, $\phi_n$ is the phase factor of propagation vector $\mathbf{Q}_{\rm 1} = (q_{\rm m}, q_{\rm m}, 1.5)$, $\mathbf{Q}_{\rm 2} = (-2q_{\rm m}, q_{\rm m}, 1.5)$, $\mathbf{Q}_{\rm 3} = (q_{\rm m}, 2q_{\rm m}, 1.5)$ with $q_{\rm m} = 0.083$. From our fitting, $\hat{e}_{\rm 1} = (0.7048, -0.4118, -0.5776)$ and $A_{\rm 1} = A_{\rm 2} = A_{\rm 3} = 0.7601, \phi_{\rm 1} = 0.1376\pi, \phi_{\rm 2} = 0.8899\pi, \phi_{\rm 3} = 0.9707\pi$. 

\begin{table}[h!]
\caption{\label{tab:tableS1} The best fitting parameter set from the Bayesian optimization and conversion into crystallographic axis. Note that the $J_{\rm c1}$ is automatically determined by $J_{\rm c1}$ = -0.031$J_{\rm c2}$. The standard deviation of the best fitting parameter set was estimated by 1$\%$ of the minimum $\chi^2$.}
    \begin{ruledtabular}
        \begin{tabular}{c c c c c c}
            (in meV) & $J_{\rm 1}$ & $K$ & $J_{\rm 3}$ & $J_{\rm c1} $& $J_{\rm c2}$ \\ \hline
            our model & -7.392 & 2.459 & 2.724 & -0.038 & 1.243 \\
            Standard deviation & 0.141 & 0.456 & 0.111 &  & 0.058 \\ \hline\hline
            (in meV) & $J_{\rm 01}$ & $\Delta_{\rm 1}$ & $J_{\rm \pm\pm}$ & $J_{\rm z\pm}$ & \\ \hline
            Crystallographic Axis & -6.6 & 1 & -0.4 & -1.13 & \\
        \end{tabular}
        \label{table:exchange set}
    \end{ruledtabular}
\end{table}

\clearpage
\begin{figure}[h]
    \includegraphics[width=0.8\columnwidth]{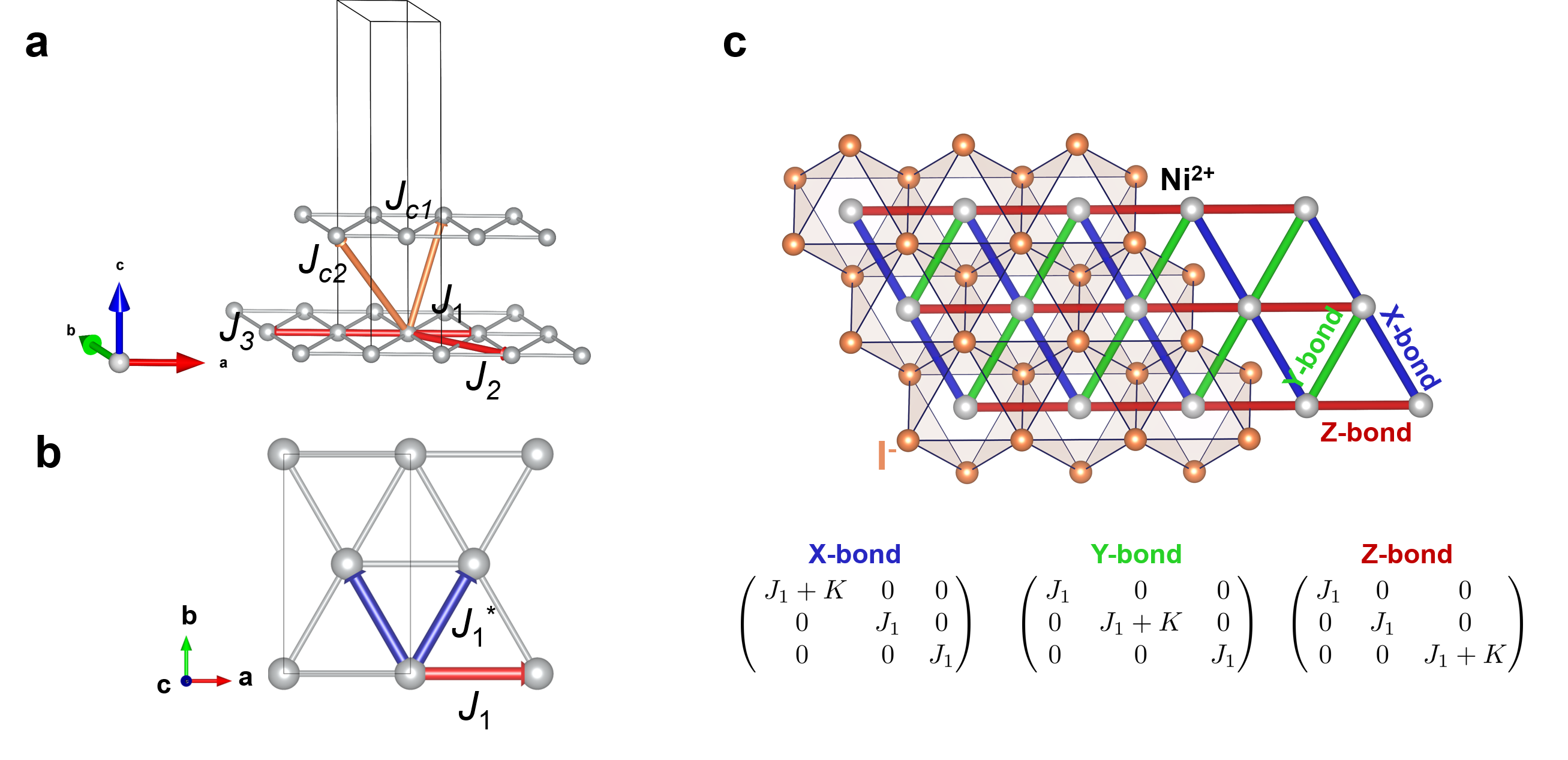}
    \centering
    \caption{\textbf{a,} Exchange path of paramagnetic $\mathrm{NiI_2}$ defined in Eq.\ref{eq:hamiltonian} in the main text. The red arrows indicate the intralayer coupling and the orange arrows indicate the interlayer coupling. \textbf{b,} Exchange path of $\mathrm{NiI_2}$ with orthorhombic unit cell. Blue and red arrows indicate the two different first nearest-neighbor interaction $J_{\rm 1}^*$ and $J_{\rm 1}$, respectively. $J_{\rm 1}^*$ is 1$\%$ smaller than the $J_{\rm 1}$ to stabilize the experimental magnetic order. \textbf{c,} Description of Kitaev-type exchange on the paramagnetic $\mathrm{NiI_2}$ lattice. Each colour bond indicates the specific Ising axis defined as a $3\times3$ matrix on the right side of the figure.}
    \label{fig:S1}
\end{figure}
\begin{figure}[h]
    \includegraphics[width=0.5\columnwidth]{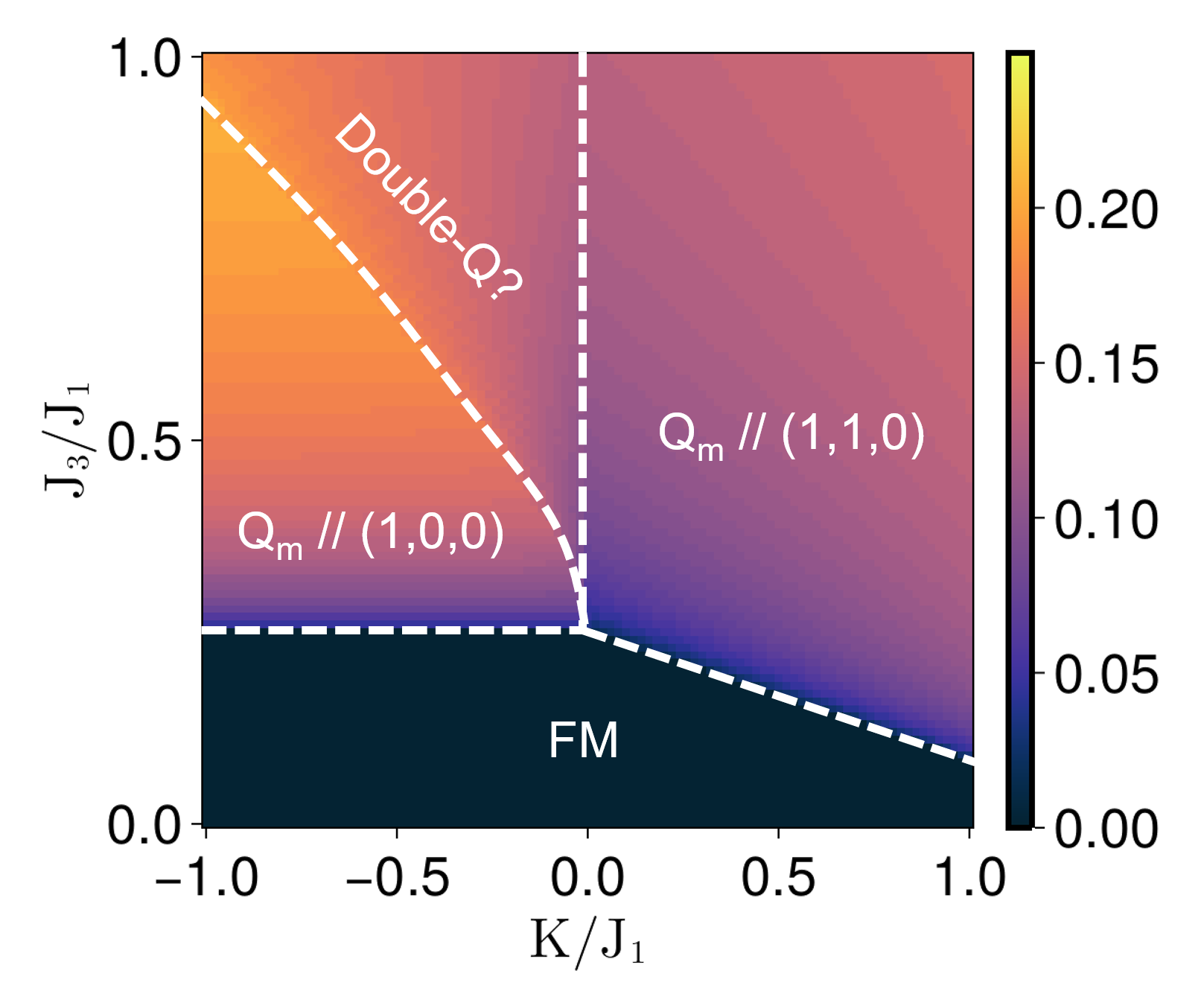}
    \centering
    \caption{Magnetic phase diagram of $J_{\rm 1}-K-J_{\rm 3}$ model using the Luttinger-Tisza method. Colourmap indicates the length of the propagation vector in units of reciprocal lattice.}
    \label{fig:S2}
\end{figure}
\begin{figure}[h]
    \includegraphics[width=0.5\columnwidth]{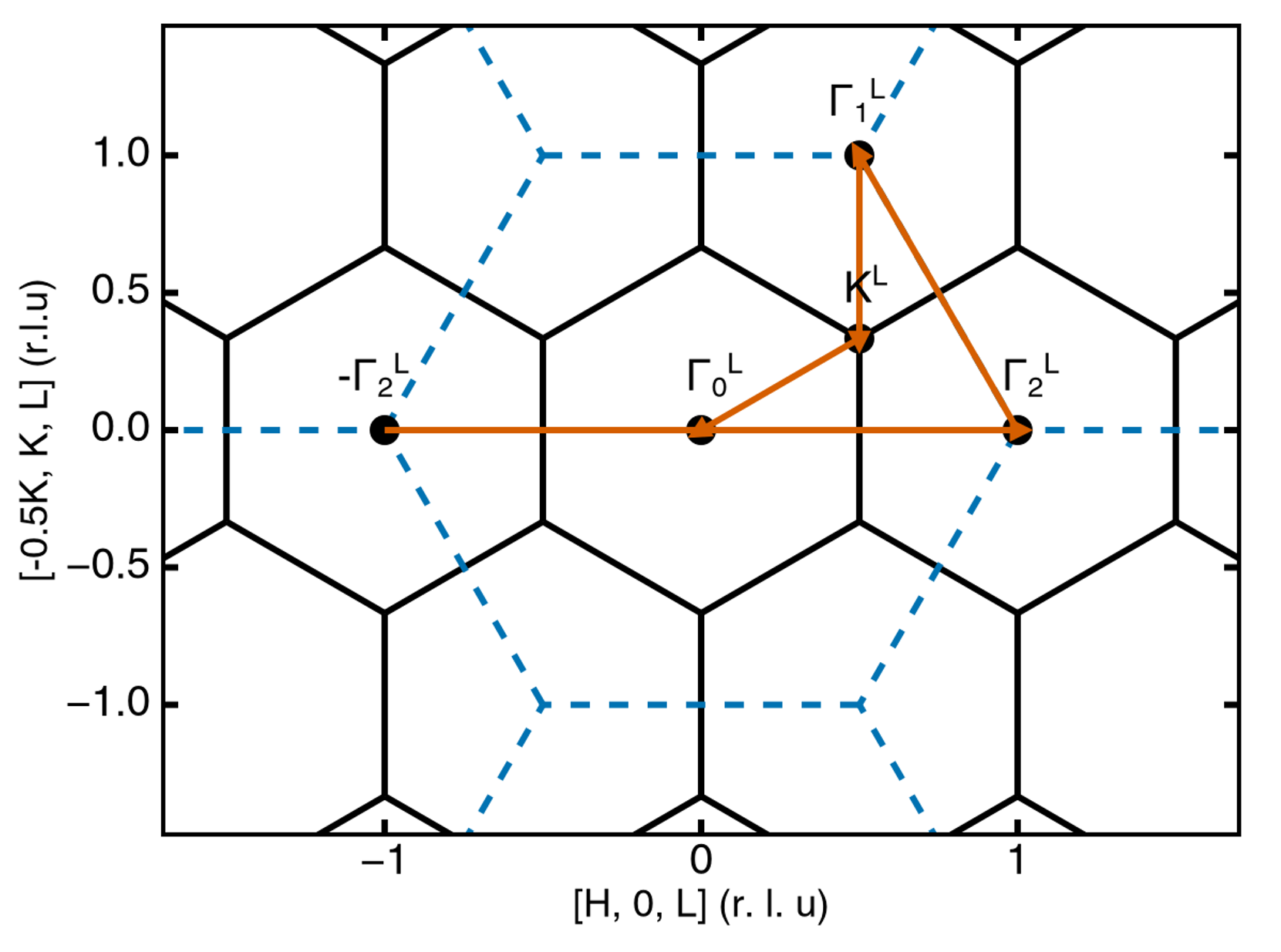}
    \centering
    \caption{Brillouin zone and convention of high-symmetric points. The orange lines indicate the path of the inelastic neutron scattering data shown in Fig.\ref{fig:2}-\ref{fig:3}.}
    \label{fig:S3}
\end{figure}
\begin{figure}[h]
    \includegraphics[width=1.0\columnwidth]{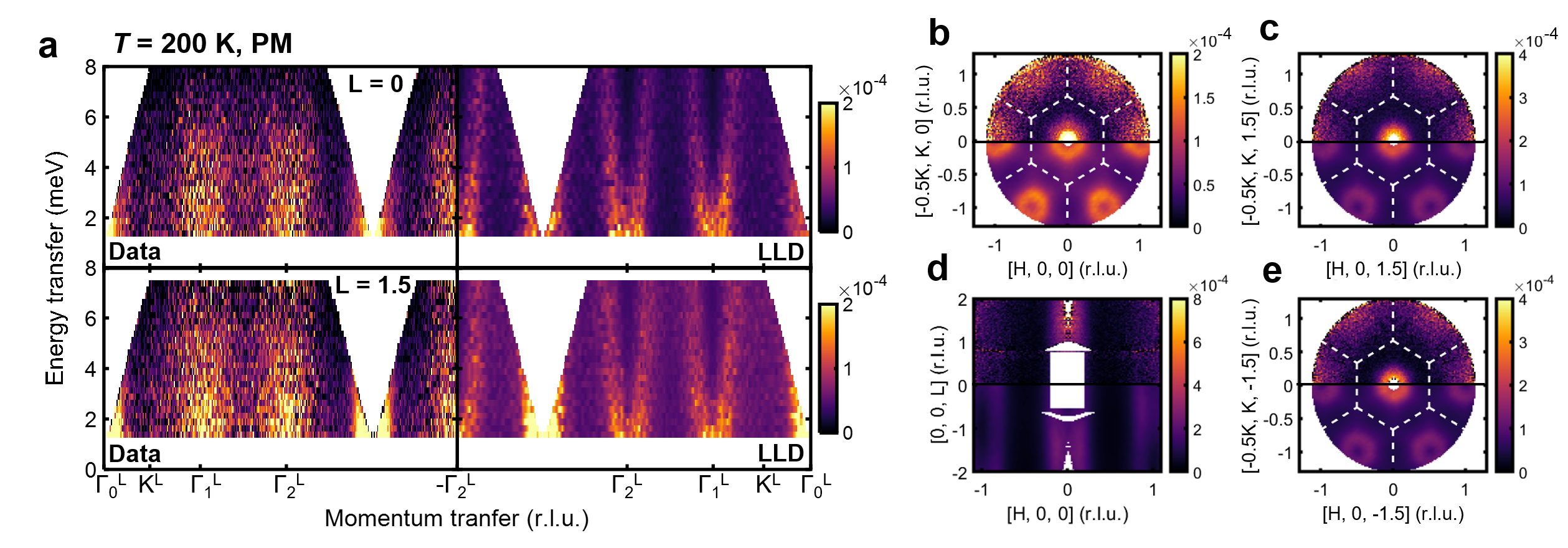}
    \centering
    \caption{\textbf{a,} Energy-resolved neutron scattering intensity in the paramagnetic regime at $T$ = 200 K along the symmetry directions of the hexagonal Brillouin zone with incident energy of $E_i$ = 11.5 meV (see Fig.\ref{fig:S2} for the directions). The upper figure is for average $\bar{L}$ = 0, and the bottom is for $\bar{L}$ = 1.5. The left side is the data, and the right side is the LLD simulation for our optimized exchange model. For the high symmetry points, $\Gamma_{0}^{L} = (0, 0, \bar{L})$, $\Gamma_{1}^{L} = (0, 1,\bar{L})$, $K^L = (1/3, 1/3, \bar{L})$, and $\Gamma_{\pm n}^{L} = (\pm n, 0, \bar{L})$. Throughout, the intensity is integrated over $\Delta L = \bar{L} + [-0.2, 0.1]$ \textbf{b-e,} Energy-integrated paramagnetic scattering intensity integrated over $\Delta E$ = [1, 6] meV from the $E_i$ = 11.5 meV data at $T$ = 100 K. \textbf{b-c, e} shows the $(H, K, \bar{L})$ plane with $\bar{L} = 0, 1.5, -1.5$, respectively. \textbf{d} shows the $(H, \bar{0}, L)$ plane integrated over $\Delta K = [-0.1, 0.1]$.}
    \label{fig:S5}
\end{figure}
\begin{figure}[h]
    \includegraphics[width=1.0\columnwidth]{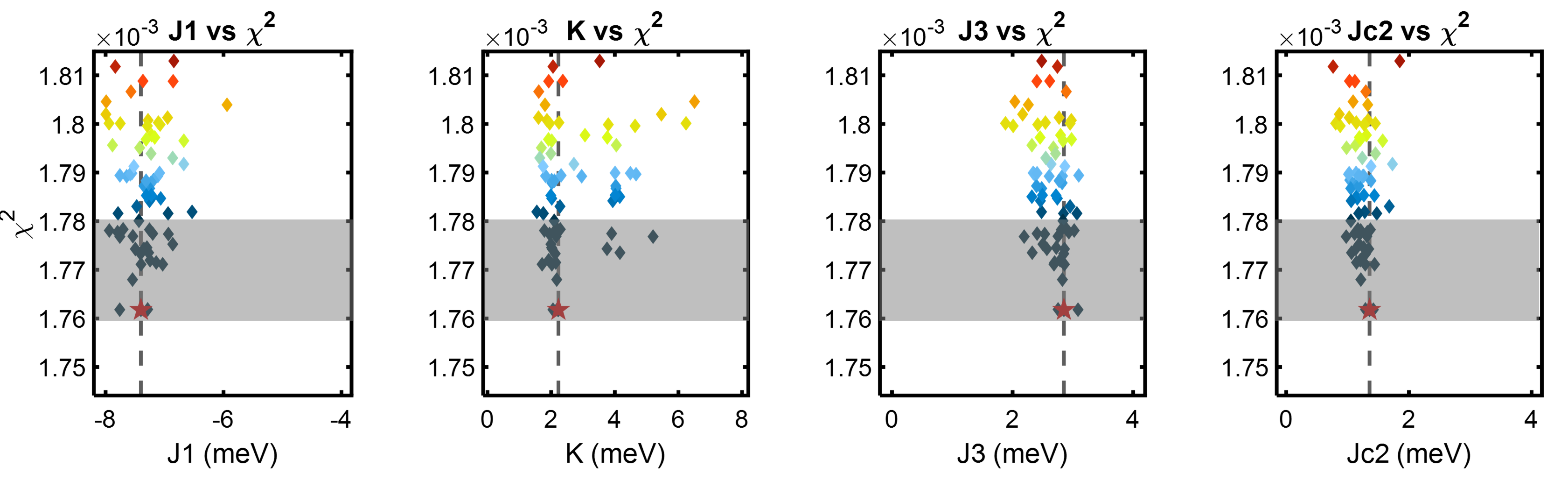}
    \centering
    \caption{The convergence of Bayesian optimization fitting for each variable. Red stars show the best-fit parameters for each exchange interaction. The grey area indicates the statistical range of calculated $\chi^2$ for the best-fit parameter set. The colour of the points indicates the value of $\chi^2$.}
    \label{fig:S6}
\end{figure}
\begin{figure}[h]
    \includegraphics[width=0.5\columnwidth]{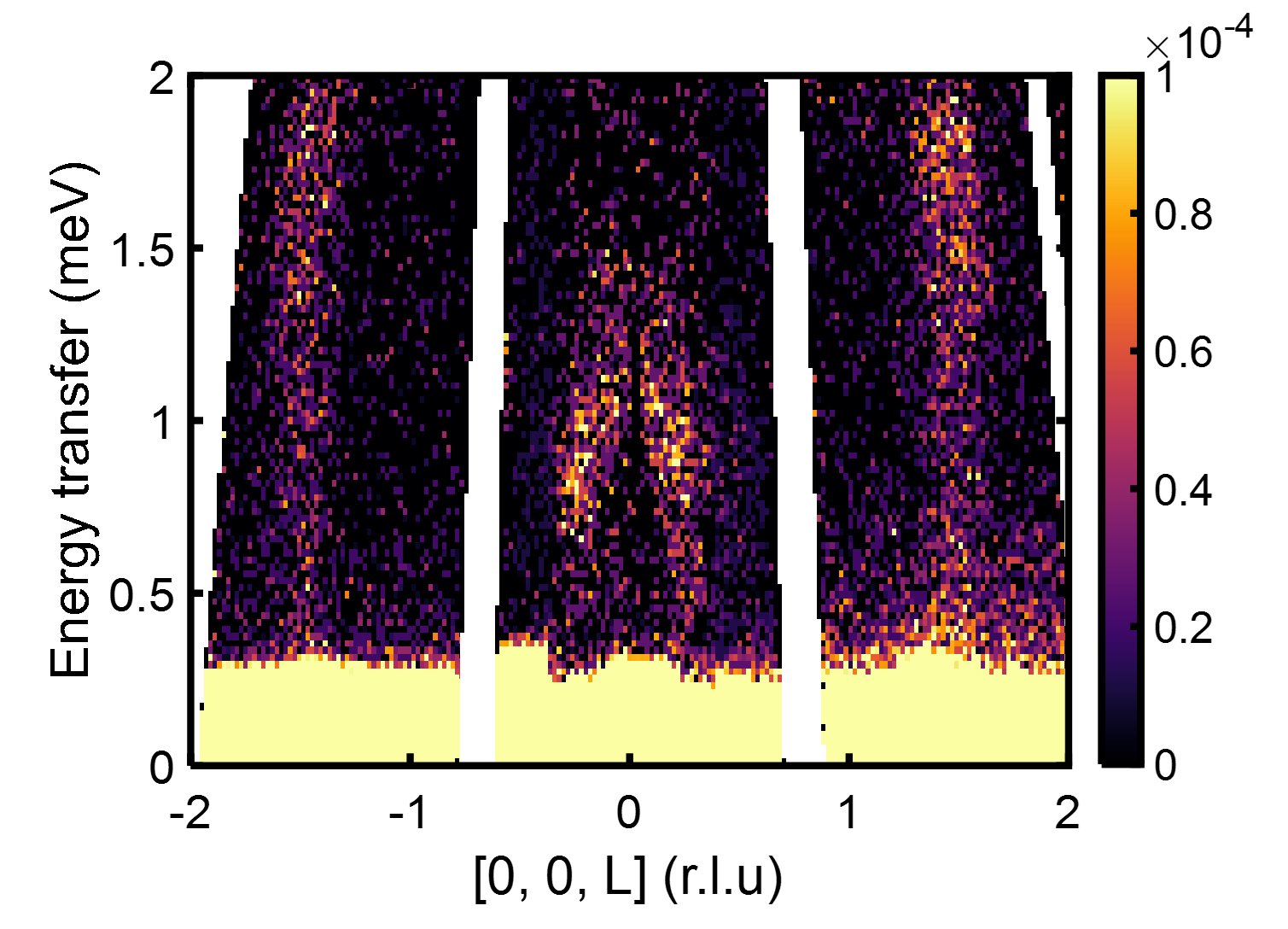}
    \centering
    \caption{Low-energy excitation of MF-Helix phase at 5 K with incident energy $E_i$ = 8 meV. The data were integrated with $\Delta H$ = [-0.2, 0.2] and $\Delta K$ = [-0.2, 0.2].}
    \label{fig:S7}
\end{figure}
\begin{figure}[h]
    \includegraphics[width=0.5\columnwidth]{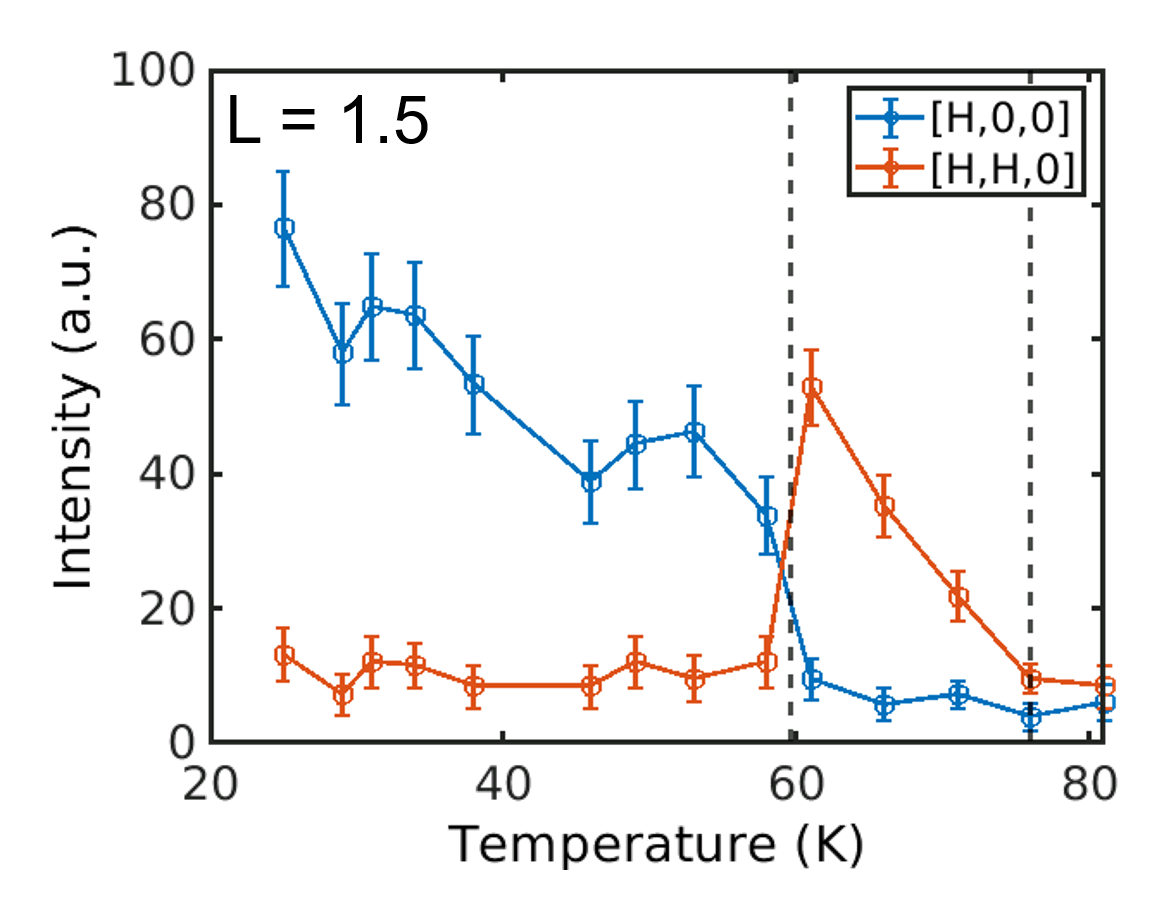}
    \centering
    \caption{Temperature dependence of the magnetic Bragg peaks measured at HRC, J-PARC, Japan. The magnetic Bragg peaks are integrated with $\Delta E$ = [-1, 1] meV and $\Delta L = 1.5+[-0.1, 0.1]$. The orange line indicates the magnetic Bragg peak along the (1, 1, 0) direction, and the blue line indicates the magnetic Bragg peak along the (1, 0, 0) direction.}
    \label{fig:S8}
\end{figure}

\begin{figure}[h]
    \includegraphics[width=0.9\columnwidth]{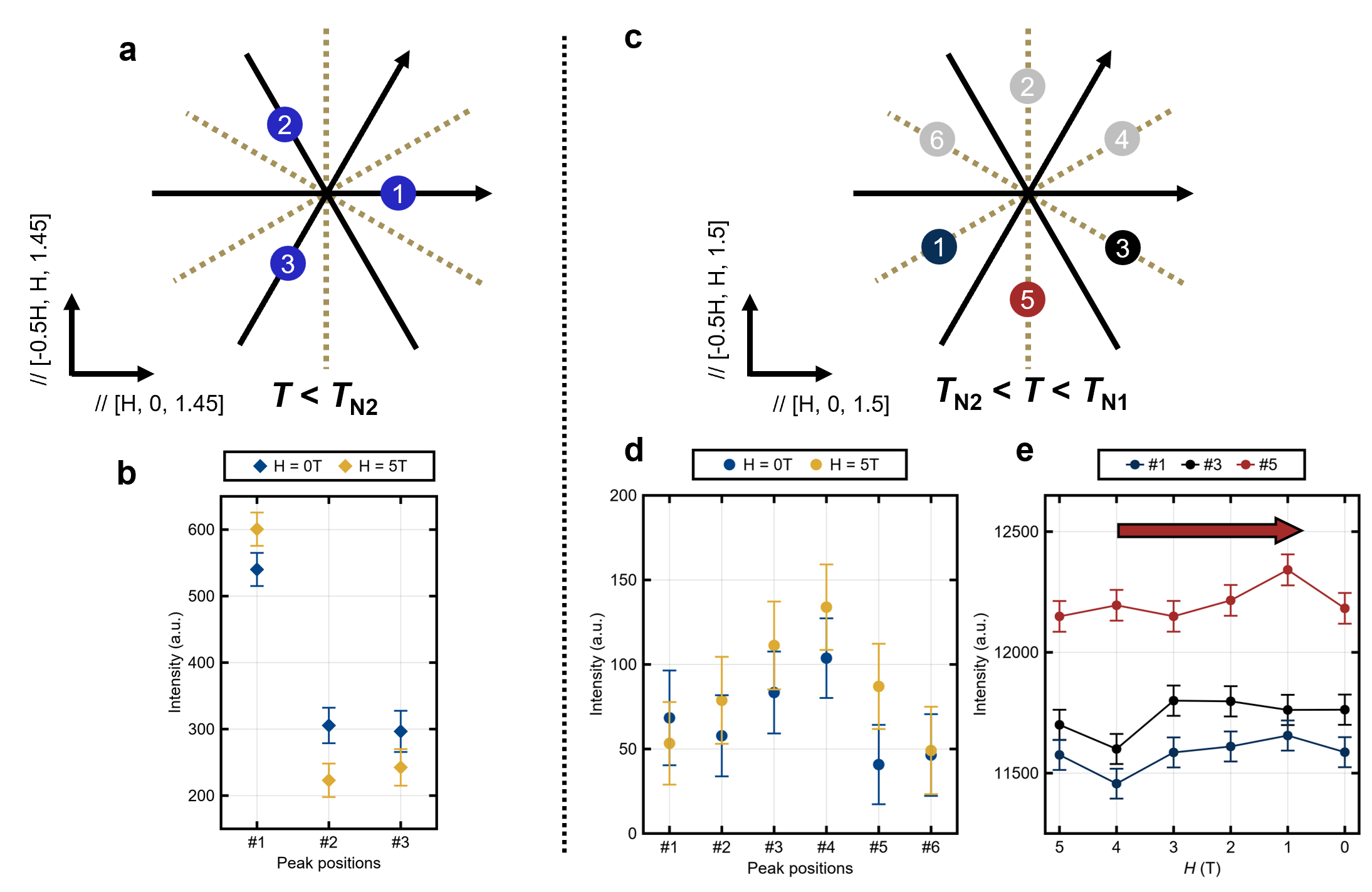}
    \centering
    \caption{\textbf{a,} Schematic view of three magnetic Bragg peaks of NiI$_2$ at $T$ = 1.5 K. Each magnetic Bragg peak is symmetrically equivalent as $\mathbf{Q_{\rm m1}} = (0.1384, 0, 1.456)$. \textbf{b,} Intensity of the three magnetic Bragg peaks after field-cooling with $\mu_0H=5$~T (gold) and turn off the magnetic field (blue). Each number corresponds to the peak in \textbf{a}. \textbf{c,} Schematic view of six magnetic Bragg peaks at $T$ = 62 K. Each magnetic Bragg peak is symmetrically equivalent as $\mathbf{Q_{\rm m2}} = (0.083, 0.083, 1.5)$. \textbf{d,} Intensity of the six magnetic Bragg peaks after field-cooling with $\mu_0H=5$~T (gold) and turn off the magnetic field (blue). Each number corresponds to the peak in \textbf{c}. \textbf{e,} Field dependence of the magnetic Bragg peak after the field training. Red arrow indicates the propagation of the field from 5T to 0T. Each color corresponds to the magnetic Bragg peak in \textbf{c} }
    \label{fig:S9}
\end{figure}


\end{document}